\documentclass[preprint,preprintnumbers, prd, floatfix, superscriptaddress,nofootinbib] {revtex4-1}
\usepackage{epsfig}
\usepackage{subfigure}
\usepackage{dcolumn}
\usepackage{bm}
\usepackage[usenames ,dvipsnames]{xcolor}
\usepackage{slashed}
\newcommand{\strich}[1]{#1  \! \! \slash}
\usepackage{graphicx,color}
\usepackage{hyperref}

\begin{document}
\title{Charmful two-body $\Omega_b$ decays in the light-front quark model}

\author{Yan-Li Wang}
\email{ylwang0726@163.com}
\affiliation{School of Physics and Information Engineering,
Shanxi Normal University, Taiyuan 030031, China}

\author{Yu-Kuo Hsiao}
\email{yukuohsiao@gmail.com}
\affiliation{School of Physics and Information Engineering,
Shanxi Normal University, Taiyuan 030031, China}

\author{Kai-Lei Wang}
\email{wangkaileicz@foxmail.com}
\affiliation{Department of Physics,
Changzhi University, Changzhi 046011, China}

\author{Chong-Chung Lih}
\email{cclih123@gmail.com}
\affiliation{School of Fundamental Physics and Mathematical Sciences,
Hangzhou Institute for Advanced Study, UCAS, Hangzhou 310024, China}

\date{\today}
\begin{abstract}
We investigate the singly and doubly charmful two-body $\Omega_b^-$ decays 
using the light-front quark model. Our findings reveal that most branching fractions 
calculated in this study, such as ${\cal B}(\Omega_b^-\to\Xi^- D^0,\Xi^{-}D^{*0})$, 
can be ten to one hundred times larger than those reported in previous calculations.
Additionally, we interpret the ratio 
${\cal B}(\Omega_b^-\to\Omega^- J/\psi)/{\cal B}(\Omega_b^-\to\Omega^- \eta_c)\simeq 3.4$ 
within the helicity framework. While the decay involving external $W$-boson emission  
appears to be suppressed by the $b\to u \bar c s$ weak transition, 
it still yields a significant branching fraction. 
For instance, 
${\cal B}(\Omega_b^-\to\Xi^0 D_s^{*-})=(5.9-14.2)\times 10^{-5}$,
${\cal B}(\Omega_b^-\to\Xi^{*0}D_s^{-})=(6.7-12.1)\times10^{-5}$, and
${\cal B}(\Omega_b^-\to\Xi^{*0}D_s^{*-})=(12.8-26.7)\times10^{-5}$,
with values reaching as large as $10^{-4}$.
These predictions are well within the experimental reach of LHCb.
\end{abstract}

\maketitle
\section{introduction}
In the $b$-baryon sextet 
${\bf B}_{6b}=(\Sigma_{b}^{(0,+,++)}$, $\Xi_b^{\prime(0,+)}$, $\Omega_b^-)$,
the predominant $\Sigma_b$ and $\Xi_b^\prime$ decays proceed via
the strong interaction, such as $\Sigma_b\to \Lambda_b^0\pi$ and $\Xi_b^\prime\to\Xi_b\pi$~\cite{pdg}. 
By contrast,  $\Omega_b^-$ cannot decay through the strong interaction and instead 
undergoes weak decays. As a result, $\Omega_b^-$ more closely resembles 
the $b$-baryon anti-triplet ${\bf B}_{3b}=(\Lambda_b^0,\Xi_b^0,\Xi_b^-)$ and
is considered the heaviest of the four weakly decaying $b$-baryon states~\cite{LHCb:2023qxn}.
Moreover, only $\Omega_b^-$, unlike $\Lambda_b^0$ or $\Xi_b$, 
can transform into the baryon decuplet~${\bf B}_{10}$,
such as $\Omega^-$ and $\Xi^{*}$, in the $b\to q$  flavor transition. 
This is because $\Omega_b^-(ssb)$, $\Omega^-(sss)$, and $\Xi^{*}(ssu,ssd)$
are associated with a symmetric strange quark pair~\cite{Gutsche:2018utw,Geng:2017mxn,
Hsiao:2020gtc,Wang:2023uea}. On the other hand, ${\bf B}_{3b}\sim (q_1 q_2-q_2 q_1)b$,
where the light quark pairs  are clearly anti-symmetric,
making the ${\bf B}_{3b}\to {\bf B}_{10}$ transition inaccessible~\cite{Hsiao:2020iwc}. Consequently, 
$\Omega_b^-$ weak decays provide a unique scenario for exploring hadronization 
and baryonic $CP$ violation~\cite{LHCb:2024exp,Geng:2005wt,
Geng:2006jt,Hsiao:2014mua,Hsiao:2019ann,Zhang:2022iye,Wang:2024rwf,
Wang:2024qff,Hsiao:2017tif,Rui:2022jff,Sinha:2021mmx}.

Of the four weakly decaying $b$-baryon states, $\Omega_b^-$ is least-well studied~\cite{LHCb:2023qxn}.
Even until very currently, $\Omega_b^-$ decays had not been extensively measured 
due to the insufficient accumulation of events. Additionally, 
the fragmentation fraction $f_{\Omega_b}$, 
which represent the $b$-quark production rate for $\Omega_b^-$, 
has yet to be firmly determined~\cite{pdg}.
It is anticipated that the doubly charmful $\Omega_b^-$ decay channels can provide
less suppressed branching fractions for observation, thereby
playing a key role in determining
the fragmentation fraction~\cite{LHCb:2023qxn,Hsiao:2015txa,Hsiao:2021mlp,Wang:2024ozz}.
For example, the LHCb collaboration has reported the partially measured branching fraction of
$\Omega_b^-\to \Omega^- J/\psi$ as well as $\Xi_b^-\to \Xi^- J/\psi$~\cite{LHCb:2023qxn}, 
given by
\begin{eqnarray}\label{data1}
{\cal R}(\Omega_b/\Xi_b)\equiv\frac{f_{\Omega_b}}{f_{\Xi_b}}\times
\frac{{\cal B}(\Omega_b^-\to \Omega^- J/\psi)}{{\cal B}(\Xi_b^-\to \Xi^- J/\psi)}
=0.120\pm 0.008\pm 0.008\,,
\end{eqnarray}
together with the most precise determination of the $\Omega_b^-$ mass to date:
$m_{\Omega_b}=(6045.9\pm 0.5\pm 0.6)$~MeV.
In Eq.~(\ref{data1}), the new data significantly deviates from the PDG result:
${\cal R}(\Omega_b/\Xi_b)=0.28\pm 0.13$~\cite{pdg}.
Utilizing the theoretical relation: ${\cal B}(\Omega_b^-\to \Omega^- J/\psi)\simeq
{\cal B}(\Xi_b^-\to \Xi^- J/\psi)$~\cite{Hsiao:2021mlp,Hsiao:2015cda},
and the new data in Eq.~(\ref{data1}),
$f_{\Omega_b}/f_{\Xi_b}\simeq 0.12$ is obtained~\cite{LHCb:2023qxn},
which aligns with theoretical estimations~\cite{Hsiao:2015txa,Hsiao:2021mlp,Wang:2024ozz}.
In fact, in heavy baryon decays, different model calculations yield varying branching fractions~\cite{Cheng:1996cs,Hsiao:2015txa,Hsiao:2015cda,Hsiao:2017tif,Hsiao:2021mlp,
Wang:2024ozz,Gutsche:2018utw,Rui:2022jff,Rui:2023fiz,
Wang:2022zja,Hsiao:2020gtc,Zhao:2018zcb},
making $f_{\Omega_b}/f_{\Xi_b}$ inconclusive.
Additionally, both $\Omega_b^-\to \Omega^- J/\psi$ and $\Xi_b^-\to \Xi^- J/\psi$
proceed with the internal $W$-boson emission topology~\cite{Hsiao:2015txa,Hsiao:2021mlp,Hsiao:2015cda},
which are sensitive to non-factorizable QCD-loop corrections. 
This raises further uncertainty in determining $f_{\Omega_b}/f_{\Xi_b}$.

To address the ambiguity in model calculations, 
additional two-body $\Omega_b^-$ decays can be investigated, such as
$\Omega_b^-\to {\bf B}^{(*)}M_{c\bar c}$, $\Omega_b^-\to {\bf B}^{(*)}M_{c}$ and
$\Omega_b^-\to {\bf B}^{(*)}\bar M_{c}$,
where $M_{c\bar c}$($M_c$) stand for a double- (single-) charm meson, and
${\bf B}^{(*)}$ denotes the octet (decuplet) baryon with spin-1/2 (3/2).
Notably, some decay channels proceed via the external $W$-boson emission configuration, 
such as $\Omega_b^-\to \Xi^{0}D_s^-$, which avoids the significant uncertainty 
associated with non-factorizable QCD-loop corrections.
Thus, we propose a systematic study of the singly and doubly charmful two-body $\Omega_b^-$ decays,
with the required form factors provided using the light-front quark model.

\section{Formalism}
%
\begin{figure}[t!]
\includegraphics[width=3.1in]{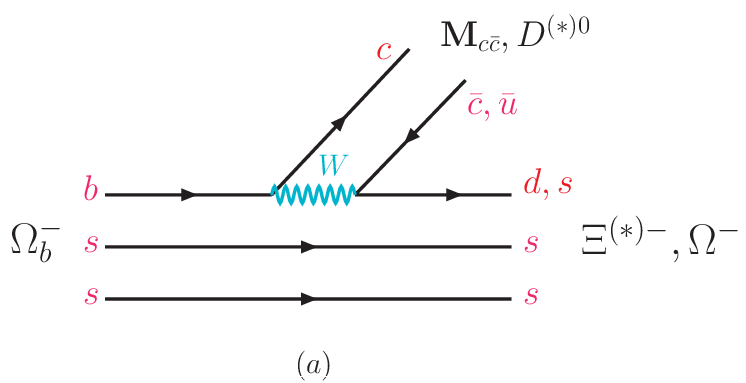}
\includegraphics[width=3.1in]{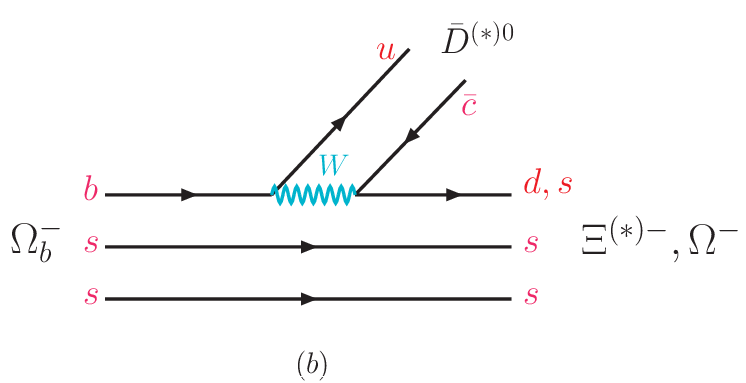}
\includegraphics[width=3in]{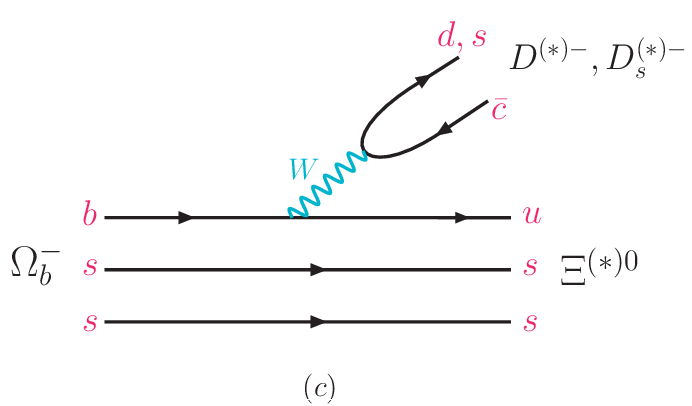}
\caption{Feynman diagrams for the charmful two-body $\Omega_b^-$ decays:
(a) $\Omega_b^-\to {\bf B}^{(*)}M_{c\bar c},{\bf B}^{(*)}D^{(*)0}$,
(b) $\Omega_b^-\to {\bf B}^{(*)}\bar D^{(*)0}$, and
(c) $\Omega_b^-\to \Xi^{(*)0}M_{\bar c}$, where $M_{c\bar c}=(\eta_c,J/\psi)$,
$M_{\bar c}=(D^{(*)-},D_s^{(*)-})$ and ${\bf B}^{(*)}=(\Xi^{(*)-},\Omega^-)$.}\label{fig1}
\end{figure}
%
We investigate the charmful two-body $\Omega_b^-$ decays, which proceed through
the quark-level weak transitions $b\to c\bar c q$, $b\to c\bar u q$, and $b\to u\bar c q$.
The corresponding effective Hamiltonian is expressed 
in a general form~\cite{Buchalla:1995vs,Buras:1998raa}
\begin{eqnarray}\label{Heff}
{\cal H}_{eff}(b\to q_1 \bar q_2 q_3)=\frac{G_F}{\sqrt 2}V_{q_1 b}V_{q_2 q_3}^*(c_1O_1+c_2O_2)+H.C.\,,
\end{eqnarray}
where $G_F$ is the Fermi constant,
$V_{q_1 b}$ and $V_{q_2 q_3}$ the Cabibbo-Kabayashi-Maskawa (CKM) matrix elements,
and $c_{1,2}$ are the scale-dependent Wilson coefficients.
The operators $O_1$ and $O_2$, which represent the quark currents, 
are defined as
\begin{eqnarray}\label{O12}
O_1=(\bar q_2 q_3)_{V-A}(\bar q_1 b)_{V-A}\,,\;
O_2=(\bar q_1 q_3)_{V-A}(\bar q_2 b)_{V-A}\,,
\end{eqnarray}
with $(\bar q_i q_j)_{V-A}\equiv \bar q_i\gamma_\mu (1-\gamma_5)q_j$
and  $(\bar q_k b)_{V-A}\equiv \bar q_k\gamma_\mu (1-\gamma_5)b$.

As a representative charmful two-body $\Omega_b^-$ decay,
$\Omega_b^-\to \Xi^- D^0$ occurs through the $\Omega_b^-\to\Xi^-$ transition
and the vacuum to $D^0$ production, as illustrated in Fig.~\ref{fig1}a.
In this process, the $\bar u$ and $d$ quarks, assigned to the $D^0$ and $\Xi^-$,
respectively, are associated with the internal $W$-boson emission. In contrast,
the $D^-$ meson in $\Omega_b^-\to \Xi^0 D^-$ is produced via
the external $W$-boson emission, as shown in Fig.~\ref{fig1}(c).
Using the effective Hamiltonian, we derive the amplitudes
as follows~\cite{Hsiao:2015txa,Hsiao:2021mlp,Hsiao:2015cda}
\begin{eqnarray}\label{amp1}
{\cal M}_1(\Omega_b^-\to {\bf B}^{(*)}M_{c\bar c})&=&
\frac{G_F}{\sqrt 2}V_{cb}V_{cq_3}^*a_2\,
\langle M_{c\bar c}|(\bar c c)_{V-A}|0\rangle
\langle {\bf B}^{(*)}|(\bar q_3 b)_{V-A}|\Omega_b^-\rangle\,,\nonumber\\
{\cal M}_2(\Omega_b^-\to {\bf B}^{(*)}D^{(*)0})&=&
\frac{G_F}{\sqrt 2}V_{cb}V_{uq_3}^*a_2\,
\langle D^{(*)0}|(\bar c u)_{V-A}|0\rangle
\langle {\bf B}^{(*)}|(\bar q_3 b)_{V-A}|\Omega_b^-\rangle\,,\nonumber\\
{\cal M}_3(\Omega_b^-\to {\bf B}^{(*)}\bar D^{(*)0})&=&
\frac{G_F}{\sqrt 2}V_{ub}V_{cq_3}^*a_2\,
\langle \bar D^{(*)0}|(\bar u c)_{V-A}|0\rangle
\langle {\bf B}^{(*)}|(\bar q_3 b)_{V-A}|\Omega_b^-\rangle\,,\nonumber\\
{\cal M}'_3(\Omega_b^-\to \Xi^{(*)0}M_{\bar c})&=&
\frac{G_F}{\sqrt 2}V_{ub}V_{cq_3}^*a_1\,
\langle M_{\bar c}|(\bar q_3 c)_{V-A}|0\rangle
\langle \Xi^{(*)0}|(\bar u b)_{V-A}|\Omega_b^-\rangle\,,
\end{eqnarray}
where $M_{c\bar c}=(\eta_c,J/\psi)$ and $q_3=(d,s)$ for
$M_{\bar c}=(D^{(*)-},D_s^{(*)-})$ and ${\bf B}^{(*)}=(\Xi^{(*)-},\Omega^-)$.
These are associated with the internal or external $W$-boson emission configuration,
as depicted in Figs.~\ref{fig1}(a, b, c). 
The parameters $a_1$ and $a_2$ result from
the factorization approximation, given by~\cite{Ali:1998eb}
\begin{eqnarray}\label{a1a2}
a_1=c_1^{eff}+\frac{c_2^{eff}}{N_c}\,, 
a_2=c_2^{eff}+\frac{c_1^{eff}}{N_c}\,,
\end{eqnarray}
where the effective Wilson coefficients $(c_1^{eff},c_2^{eff})$
account for the vertex and penguin corrections to $(c_1,c_2)$, 
and $N_c$ represents the color number.

Actually, 
the factorization leads to $a_2=c_2^{eff}+c_1^{eff}/N_c+c_1 \langle \chi_1\rangle$
for the amplitude ${\cal M}_1(\Omega_b^-\to {\bf B}^{(*)}M_{c\bar c})$.
The additional term, 
\begin{eqnarray}
\langle \chi_1\rangle=\frac{
\langle {\bf B}^{(*)}M_{c\bar c}|2\bar c\gamma_\mu(1-\gamma_5)T^a c
\bar q_3\gamma^\mu(1-\gamma_5)T^a b|\Omega_b^-\rangle}
{\langle M_{c\bar c}|(\bar c c)_{V-A}|0\rangle\langle {\bf B}^{(*)}|(\bar q_3 b)_{V-A}|\Omega_b^-\rangle}\,,
\end{eqnarray}
is regarded as a non-factorizable QCD correction. This is because
the color operators $T^a$ induce gluon exchange between the two currents, 
resulting in an inseparable connection between the final states. 
For the amplitude ${\cal M}'_3(\Omega_b^-\to \Xi^{(*)0}M_{\bar c})$, 
a similar non-factorizable QCD correction appears as $\langle \chi_2\rangle$ in
$a_1=c_1^{eff}+c_2^{eff}/N_c+c_2 \langle \chi_2\rangle$. We conclude that
the terms involving $\langle \chi_1\rangle$ and $\langle \chi_2\rangle$ 
are typically present in tree-dominated decays. In the naive factorization approach, 
these non-factorizable terms are neglected, resulting in their absence from Eq.~(\ref{a1a2}).
However, this simplification is disfavored by experimental data.
Therefore, the generalized factorization approach is adopted~\cite{Ali:1998eb},
where the effective color number $N_c^{eff}$ replaces the naive one in Eq.~(\ref{a1a2}). 
Empirically, varying $N_c^{eff}$ from 2, 3, to $\infty$ provides an estimation of the impacts 
from non-factorizable QCD corrections~\cite{Ali:1998eb,Geng:2005fh,Geng:2007cw}.

In Eq.~(\ref{amp1}), the matrix elements of the vacuum to meson production
are parameterized as~\cite{Hsiao:2015cda,Chen:2008sw,pdg}
\begin{eqnarray}\label{dconst}
\langle P|(\bar q' q)_{V-A}|0\rangle=if_P p_\mu\,,\;
\langle V|(\bar q' q)_{V-A}|0\rangle=m_V f_V \epsilon_\mu\,,
\end{eqnarray}
where $P(V)$ denotes the pseudo-scalar (vector) meson,
and $f_{P(V)}$ represents the decay constant. The matrix elements 
for the $\Omega_b^-\to {\bf B}^{(*)}$ transition
are expressed as~\cite{Gutsche:2018utw,Ke:2007tg,Zhao:2018zcb}
\begin{eqnarray}\label{FF1}
&&
\langle T^{\mu}_{\frac{1}{2}} \rangle_{V-A}
\equiv \langle{\bf B}(P',S'=
\frac{1}{2},S_{z}^{\prime})|(\bar q b)_{V-A}|\Omega_b(P,S=\frac{1}{2},S_{z})\rangle \nonumber\\
&=& \bar{u}(P',S'_{z})
\left[\gamma^{\mu} f_{1}^V
+i \frac{f_{2}^V}{M}\sigma^{\mu\nu}p_{\nu}
+\frac{f_{3}^V}{M} p^{\mu}\right]
u(P,S_{z})\nonumber\\
&&-\bar{u}(P',S_{z}^{\prime})
\left[\gamma^{\mu} f_{1}^A
+i \frac{f_{2}^A}{M}\sigma^{\mu\nu}p_{\nu}
+\frac{f_{3}^A}{M} p^{\mu}\right]\gamma_{5}u(P,S_{z})\,,
\end{eqnarray}
and~\cite{Zhao:2018mrg,Gutsche:2018utw,Hsiao:2021mlp,Hsiao:2020gtc}
\begin{eqnarray}\label{FF2}
&&\langle T_{3\over 2}^\mu\rangle_{V-A}
\equiv
\langle {\bf B}^*(P^{\,\prime},S'={3\over 2},S_z^\prime)|(\bar q b)_{V-A}
|\Omega_b(P,S={1\over 2},S_z)\rangle \nonumber\\
&& =  \bar{u}_{\alpha}(P^{\,\prime},S_{z}^{\,\prime})
\left[\frac{P^{\alpha}}{M}\left(\gamma^{\mu}F^V_{1}
+\frac{P^{\mu}}{M} F^V_{2}
+\frac{P^{\,\prime\mu}}{M'}F^V_{3}\right)+g^{\alpha\mu}F^V_{4}\right]
\gamma_{5}u(P,S_{z})\nonumber\\
& &\quad-\bar{u}_{\alpha}(P^{\,\prime},S_{z}^{\,\prime})
\left[\frac{P^{\alpha}}{M}\left(\gamma^{\mu}F^A_{1}
+\frac{P^{\mu} }{M}F^A_{2}
+\frac{P^{\,\prime\mu}}{M'}F^A_{3}\right)
+g^{\alpha\mu}F^A_{4}\right]u(P,S_{z})\,,
\end{eqnarray}
where $p_\mu=(P-P')_\mu$, and $(M,M')$
represent the masses of $(\Omega_b^-,{\bf B}^{(*)})$.
The parameters $f^{V,A}_i$ ($i=1,2,3$) and $F^{V,A}_j$ ($j=1,2, ..,4$)
are corresponding form factors.

The matrix elements in Eqs.~(\ref{FF1}, \ref{FF2}) can be utilized to
calculate the helicity amplitudes, 
defined as~\cite{Gutsche:2018utw,Korner:1992wi}
\begin{eqnarray}\label{helicityA}
H^{V(A)}_{\lambda_{\bf B(B^*)} \lambda_M}
\equiv\langle {\bf B}({\bf B}^{*})|(\bar q b)_{V(A)}|\Omega_b^-\rangle
\varepsilon^\mu_M\,,
\end{eqnarray}
where $\varepsilon^\mu_M=(p^\mu/\sqrt{p^2},\epsilon_\lambda^{\mu})$ with $M=(P,V)$
follows from Eq.~(\ref{dconst}) for meson production. Here,
$\lambda_{\bf B}=\pm 1/2$ and $\lambda_{{\bf B}^*}=(\pm 3/2,\pm 1/2)$
denote the helicity states of the spin-1/2 and 3/2 baryons, respectively. 
The zero helicity state of a pseudoscalar meson is represented as $\lambda_P=\bar 0$,
while the three helicity states of a vector meson are given by $\lambda_V=(1,0,-1)$.
According to helicity conservation, the relation
$\lambda_{\Omega_b}=\lambda_{{\bf B}({\bf B}^*)}-\lambda_M$ holds,
where $\lambda_{\Omega_b}=\pm 1/2$.

With the matrix elements substituted from Eq.~(\ref{FF1}),
the helicity amplitudes $H^{V(A)}_{\lambda_{{\bf B}} \lambda_M}$ are expended as
\begin{eqnarray}\label{HVA1}
&&
H^{V(A)}_{{1\over 2}\bar 0}=\sqrt{\frac{Q_\pm^2}{p^2}}
\bigg(f_1^{V(A)}M_{\mp}\pm f_3^{V(A)}\frac{p^2}{M} \bigg)\,,
\end{eqnarray}
for $\Omega_b^-\to {\bf B}P$, and
\begin{eqnarray}\label{HVA1b}
&&
H^{V(A)}_{{1\over 2}0}=\sqrt{\frac{Q_\mp^2}{p^2}}
\bigg(f_1^{V(A)}M_{\pm}\pm f_2^{V(A)}\frac{p^2}{M}\bigg)\,,\;\nonumber\\
&&
H^{V(A)}_{{1\over 2}1}=-\sqrt{2Q_\mp^2}
\bigg(f_1^{V(A)}\pm f_2^{V(A)}\frac{M_\pm}{M} \bigg)\,,
\end{eqnarray}
for $\Omega_b^-\to{\bf B}V$, where
$M_\pm = M\pm M'$, $Q_\pm^2= M_\pm^2 - p^2$, and
$H^{V(A)}_{-\lambda_{\bf B} -\lambda_M} = \pm H^{V(A)}_{\lambda_{\bf B}\lambda_M}$.
Similarly, $H^{V(A)}_{\lambda_{{\bf B}^*} \lambda_M}$ are given as
\begin{eqnarray}\label{HVA2}
H_{\frac12 {\bar 0}}^{*V(A)} &=&\sqrt{\frac{2}{3}\frac{Q^2_{\pm}}{p^2}}
\left(\frac{Q^2_\mp}{2MM'}\right)
(F_1^{V(A)} M_\pm \mp  F_2^{V(A)}\bar M_+ \mp  F_3^{V(A)}\bar M'_- \mp F_4^{V(A)} M )\,,
\end{eqnarray}
for $\Omega_b^-\to {\bf B}^*P$, and
\begin{eqnarray}\label{HVA3}
&&
H_{\frac120}^{*V(A)}= \sqrt{\frac{2}{3}\frac{Q^2_\mp}{p^2}}
\left[ F_1^{V(A)} \left(\frac{Q^2_\pm M_\mp}{2MM'}\right)
\mp\left(F_2^{V(A)}+F_3^{V(A)}\frac{M}{M'}\right) \left(\frac{|\vec{P}'|^2}{M'}\right)
\mp F_4^{V(A)}\bar M'_- \right]\,,\nonumber\\
&&
H_{\frac121}^{*V(A)}=-\sqrt{\frac{Q^2_\mp}{3}}
\left[F_1^{V(A)} \left(\frac{Q^2_\pm}{M M'}\right) -F_4^{V(A)}\right]\,,\nonumber\\
&&
H_{\frac321}^{*V(A)} = \mp \sqrt{Q^2_\mp} \, F_4^{V(A)}\,,
\end{eqnarray}
for $\Omega_b^-\to {\bf B}^*V$, where
$|\vec{P}'|=\sqrt{Q^2_+ Q^2_-}/(2M)$,
$\bar M_{\pm}^{(\prime)}=(M_+M_-\pm p^2)/(2M^{(\prime)})$, and
$H^{*V(A)}_{-\lambda_{\bf B^*} -\lambda_M}
=\mp H^{*V(A)}_{\lambda_{\bf B^*}\lambda_M}$.

The amplitudes of $\Omega_b^-\to {\bf B}^{(*)}M$
can be expressed in the helicity basis~\cite{Gutsche:2018utw,Korner:1992wi}.
For example, we present ${\cal M}(\Omega_b^-\to \Xi^{(*)-}M_{c\bar c})$ as
\begin{eqnarray}
{\cal M}(\Omega_b^-\to \Xi^{(*)-}M_{c\bar c})=
\sum_{\lambda_{{\bf B(B^*)}}\lambda_M}
C_{\rm w} H_{\lambda_{{\bf B (B^*)}}\lambda_M}\,,
\end{eqnarray}
where $\sqrt 2 C_{\rm w}=i G_FV_{cb}V^*_{cd}a_2f_{\eta_c}m_{\eta_c}$ for $M_{c\bar c}=\eta_c$ and
$\sqrt 2 C_{\rm w}=G_F V_{cb}V^*_{cd}a_2f_{J/\psi}m_{J/\psi}$ for $M_{c\bar c}=J/\psi$. Here,
$H^{(*)}_{\lambda_{\bf B(B^*)}\lambda_M}=
H^{(*)V}_{\lambda_{\bf B(B^*)}\lambda_M}-
H^{(*)A}_{\lambda_{\bf B(B^*)}\lambda_M}$
with $H^{(*)V}$ and $H^{(*)A}$ calculated using
Eqs.~(\ref{HVA1}), (\ref{HVA1b}, (\ref{HVA2}) and (\ref{HVA3}).
The branching fractions, averaged over the spin of the $\Omega_b$ baryon,
are given by~\cite{Hsiao:2020gtc,Gutsche:2018utw}
\begin{eqnarray}\label{amp2a}
{\cal B}(\Omega^-_b\to \Xi^{(*)-} M_{c\bar c})&=&
\frac{\tau_{\Omega_b}|\vec{P}'|}{16\pi m_{\Omega_b}^2}
C_{\rm w}^2 H_{{\bf B(B^*)}M}^{2}\,,
\end{eqnarray}
where
\begin{eqnarray}\label{amp2b}
H_{{\bf B}P}^2&=&
\left|H_{\frac12{\bar 0}}\right|^2+\left|H_{{-\frac12}{\bar 0}}\right|^2\,,\nonumber\\
H_{{\bf B}V}^2&=&
\left|H_{\frac121}\right|^2+\left|H_{\frac120}\right|^2
+\left|H_{-\frac120}\right|^2+\left|H_{-\frac12-1}\right|^2\,,
\end{eqnarray}
for $\Omega^-_b\to \Xi^{-} \eta_c$ and $\Omega^-_b\to \Xi^{-}J/\psi$, and
\begin{eqnarray}\label{amp2c}
H_{{\bf B}^* P}^2&=&
\left|H^*_{\frac12{\bar 0}}\right|^2+\left|H^*_{{-\frac12}{\bar 0}}\right|^2\,,\nonumber\\
H_{{\bf B}^*V}^2&=&
\left|H^*_{\frac321}\right|^2+\left|H^*_{\frac121}\right|^2+\left|H^*_{\frac120}\right|^2
+\left|H^*_{-\frac120}\right|^2+\left|H^*_{-\frac12-1}\right|^2+\left|H^*_{-\frac32-1}\right|^2\,,
\end{eqnarray}
for $\Omega^-_b\to \Xi^{*-} \eta_c$ and $\Omega^-_b\to \Xi^{*-}J/\psi$, respectively.
The branching fractions for other decays can be determined similarly using the equations provided.

Either $\Omega_b^-$ or ${\bf B}^{(*)}$
can be viewed as a bound state of three quarks $q_1$, $q_2$ and $q_3$.
Utilizing the light front quark model, the baryon bound state is expressed as~\cite{Dosch:1988hu}
\begin{eqnarray}\label{boundstate}
|{\bf B}(P,S,S_{z})\rangle & = & \int\{d^{3}p_{1}\}
\{d^{3}p_{2}\}2(2\pi)^{3}\delta^{3}(\tilde{P}-\tilde{p}_{1}-\tilde{p}_{2})\nonumber \\
&  & \times\sum_{\lambda_{1},\lambda_{2}}\Psi^{SS_{z}}
(\tilde{p}_{1},\tilde{p}_{2},\lambda_{1},\lambda_{2})|q_{1}(p_{1},\lambda_{1})q_{[2,3]}
(p_{2},\lambda_{2})\rangle\,,
\end{eqnarray}
where $q_2$ and $q_3$ combine as a diquark, denoted by $q_{[2,3]}$.
Here, $(p_1,\lambda_1)$ and $(p_2,\lambda_2)$ represent
the momentum and helicity of $q_1$ and $q_{[2,3]}$, respectively,
while $\Psi^{SS_{z}}$ is the wave function.
In the light-front framework, we write $P=(P^-,P^+,P_\bot)$
with $P^\pm=P^0\pm P^3$ and $P_\bot=(P^1,P^2)$.
Similarly, $p_i=(p_i^-,p_i^+,p_{i\bot})$, where $i=$1 or 2, and
$p_i^\pm=p_i^0\pm p_i^3$, $p_{i\bot}= (p_i^1, p_i^2)$.
Using the identities $P^+ P^-=M^2+P_{\bot}^2$ and $p_i^+ p_i^- = {m_i^2+p_{i\bot}^2}$,
we reduce $P^-$ and $p_i^-$, where $(m_1,m_2)=(m_{q_1},m_{q_2}+m_{q_3})$.
Thus, $\tilde P\equiv(P^+,P_\bot)$ and $\tilde p_i\equiv(p_i^+, p_{i\bot})$
are applied in the light-front quark model,
with the relations
\begin{eqnarray}\label{LF2}
&&
p^+_1=(1-x) P^+\,,\;
p^+_2=x P^+\,,\;\nonumber\\
&&
p_{1\bot}=(1-x) P_\bot-k_\bot\,,\;
p_{2\bot}=xP_\bot+k_\bot\,,
\end{eqnarray}
where $k_\perp$ is the relative momentum from $\vec{k}=(k_\perp,k_z)$,
ensuring $P^{+}=p^+_1+p^+_2$ and $P_{\bot}=p_{1\bot}+p_{2\bot}$.

By performing the Melosh transformation~\cite{Melosh:1974cu},
where $e_{1,2}\equiv\sqrt{m^2_{1,2}+\vec{k}^2}$ and $M_0\equiv e_1+e_2$,
we obtain
\begin{eqnarray}\label{Melosh1}
&&
x=\frac{e_2-k_z}{e_1+e_2}\,,\quad 1-x=\frac{e_1+k_z}{e_1+e_2}\,,\quad
k_z=\frac{xM_0}{2}-\frac{m^2_{2}+k^2_{\perp}}{2xM_0}\,,\nonumber\\
&&
M_0^2={ m_{1}^2+k_\bot^2\over 1-x}+{ m_{2}^2+k_\bot^2\over  x}\,,
\end{eqnarray}
with $x$, $1-x$, and $k_z$ being rewritten. Particularly,
$(\bar P_\mu \gamma^\mu-M_0)u(\bar{P},S_{z})=0$ is obtained,
where $\bar P\equiv p_1+p_2$ illustrates the internal motions
of $q_1$ and $q_{[2,3]}$ within the baryon~\cite{Jaus:1991cy}.
Consequently, the wave function is 
derived as~\cite{Ke:2012wa,Ke:2017eqo,Zhao:2018mrg,Hu:2020mxk,Ke:2019smy}
\begin{eqnarray}\label{WF}
\Psi^{SS_{z}}(\tilde{p}_{1},\tilde{p}_{2},\lambda_{1},\lambda_{2})=
\frac{C_{S,A}^{(\alpha)}}{\sqrt{2(p_{1}\cdot\bar{P}+m_{1}M_{0})}}\bar{u}(p_{1},\lambda_{1})
\Gamma_{S,A}^{(\alpha)} u(\bar{P},S_{z})\phi(x,k_{\perp})\,,
\end{eqnarray}
with the subscripts $S$ and $A$ denoting 
the scalar and axial-vector combinations of the diquarks.
As the quark-diquark coupling vertex functions in the baryon state,
$\Gamma_{S,A}^{(\alpha)}$ are derived as~\cite{Cheng:2004cc}
\begin{eqnarray}
\Gamma_{S}=1\,,\;
\Gamma_{A} =-\frac{1}{\sqrt{3}}\gamma_{5} \strich\epsilon^{*}(p_{2},\lambda_{2})\,,\;
\Gamma_{A}^{\alpha} =\epsilon^{*\alpha}(p_{2},\lambda_{2})\,,
\end{eqnarray}
together with $C_{S,A}^{(\alpha)}$ the normalization functions,
given by
\begin{eqnarray}
(C_{S(A)},C_{A}^{\alpha})=\bigg(
\frac{3(m_{1}M_{0}+p_{1}\cdot\bar{P})}{3m_{1}M_{0}+p_{1}\cdot\bar{P}+
2(p_{1}\cdot p_{2})(p_{2}\cdot\bar{P})/m_{2}^{2}}\,,\;
\frac{3m_{2}^{2}M_{0}^{2}}{2m_{2}^{2}M_{0}^{2}+(p_{2}\cdot\bar{P})^{2}}
\bigg)\,.
\end{eqnarray}
In Eq.~(\ref{WF}), $\phi(x,k_{\perp})$ describes the momentum distribution
of constituent quarks $q_1$ and $q_{[2,3]}$ in the bound state,
presented in Gaussian form~\cite{Hsiao:2020gtc,Zhao:2018zcb,Ke:2012wa,
Ke:2017eqo,Zhao:2018mrg,Hu:2020mxk,Ke:2019smy}
\begin{eqnarray}
\phi(x,k_{\perp})=4\left(\frac{\pi}{\beta^{2}}\right)^{3/4}\sqrt{\frac{dk_{z}}{dx}}\exp
\left(\frac{-\vec{k}^{2}}{2\beta^{2}}\right)\,,
\label{DA}
\end{eqnarray}
where $k_z$ can be found in Eq.~(\ref{Melosh1}), and
$\beta\equiv\beta_{q_1[q_2 q_3]}$ shapes the momentum distribution
of the $q_1$-$q_{[2,3]}$ baryon bound state.

With $|\Omega_b(P,S,S_z)\rangle$
and $|{\bf B}^{(*)}(P',S',S'_z)\rangle$ presented with Eq.~(\ref{boundstate}),
the matrix elements of the $\Omega_b\to {\bf B}^{(*)}$ transition
can be derived in the light-front framework,
given by~\cite{Ke:2007tg,Zhao:2018zcb}:
\begin{eqnarray}\label{transitionVA2}
\langle \hat{T}^{\mu}_{\frac{1}{2}} \rangle_{V-A}
&\equiv& \langle {\bf B}(P',S'=1/2,S_z^\prime)|(\bar q b)_{V-A}|\Omega_b(P,S=1/2,S_z)\rangle\nonumber \\
&=&
N_{FS}\int\{d^{3}p_{2}\}\hat C^{-1/2} \phi^{\prime}(x^{\prime},k_{\perp}^{\prime})\phi(x,k_{\perp})
\nonumber \\
&& 
\times\sum_{\lambda_{2}}\bar{u}(\bar{P}',S'_{z})
\left[\bar{\Gamma}^{\prime}_{S(A)}(\strich p_{1}^{\prime}+m_{1}^{\prime})
\gamma^{\mu}(1-\gamma_{5})(\strich p_{1}+m_{1})\Gamma_{S(A)}\right]u(\bar{P},S_{z})\,.
\end{eqnarray}
And~\cite{Zhao:2018mrg,Hsiao:2020gtc,Hsiao:2021mlp}:
\begin{eqnarray}\label{transition31}
\langle \hat{T}^{\mu}_{\frac{3}{2}} \rangle_{V-A}
&\equiv&
\langle {\bf B}^*(P',S'=3/2,S'_z)|(\bar q b)_{V-A}|\Omega_b(P,S=1/2,S_z)\rangle\nonumber \\
&=&
N_{FS}\int\{d^{3}p_{2}\}\hat C^{-1/2} \phi^{\prime}(x^{\prime},k_{\perp}^{\prime})\phi(x,k_{\perp})
\nonumber \\
&& 
\times\sum_{\lambda_{2}}\bar{u}_{\alpha}(\bar{P}',S'_{z})
\left[\bar{\Gamma}^{\prime\alpha}_{A}(\strich p_{1}^{\prime}+m_{1}^{\prime})
\gamma^{\mu}(1-\gamma_{5})(\strich p_{1}+m_{1})\Gamma_{A}\right]u(\bar{P},S_{z})\,,
\end{eqnarray}
where $\bar \Gamma=\gamma^0 \Gamma^\dagger\gamma^0$,
$\hat C=4p_{1}^{+}p_{1}^{\prime+}(p_{1}\cdot\bar{P}
+m_{1}M_{0})(p_{1}^{\prime}\cdot\bar{P}'+m_{1}^{\prime}M_{0}^{\prime})$,
and $N_{FS}$ is a flavor-spin factor varying in different processes.
According to
$I^{\mu}_{(5)}\equiv\bar{u}\Gamma^{\mu}_{1/2}(\gamma_5)u$,
$\hat I^{\mu}_{(5)}\equiv\bar{u}\hat \Gamma^{\mu}_{1/2}(\gamma_5)u$,
$J_{(5)}^{\mu}\equiv\bar{u}\Gamma^{\mu\beta}_{3/2}(\gamma_5) u_{\beta}$, and
$\hat J_{(5)}^{\mu}\equiv\bar{u}\hat \Gamma^{\mu\beta}_{3/2}(\gamma_5) u_{\beta}$,
where
\begin{eqnarray}
&&
(\Gamma^{\mu}_{1\over 2})_i=
(\gamma^{\mu}, P^{\mu}, P^{\prime\mu})\,,\;
\nonumber\\&&
(\hat \Gamma^{\mu}_{1\over 2})_i=
(\gamma^{\mu}, \bar{P}^{\mu}, \bar{P}^{\prime\mu})\,,\nonumber\\
&&
(\Gamma^{\mu\beta}_{3\over 2})_j=
(\gamma^{\mu}P^{\beta},P^{\,\prime\mu}P^{\beta},P^{\mu}P^{\beta},g^{\mu\beta})\,,\;
\nonumber\\&&
(\hat \Gamma^{\mu\beta}_{3\over 2})_j=
(\gamma^{\mu}\bar P^{\beta},\bar P^{\,\prime\mu}\bar P^{\beta},\bar P^{\mu}\bar P^{\beta},g^{\mu\beta})\,,
\end{eqnarray}
with $i=1,2,3$ and $j$=1, 2, 3, ..., 4,
we perform the following projections~\cite{Ke:2007tg,Zhao:2018zcb,Zhao:2018mrg,Hsiao:2020gtc}:
\begin{eqnarray}\label{F5j}
&&
(I^{\mu})_i\cdot \langle T_{1\over 2}\rangle_V={\rm Tr}\{u(P',S'_{z})\bar{u}(P',S'_{z})
\left[\gamma^{\mu} f_{1}^V
+i \frac{f_{2}^V}{M}\sigma^{\mu\nu}p_{\nu}
+\frac{f_{3}^V}{M'} p^{\mu}\right]u(P,S_{z})
\bar{u}(P,S_{z})({\Gamma}^{\mu}_{1\over 2})_i\}\,,
\nonumber\\
&&
(\hat I^{\mu})_i\cdot \langle \hat T_{1\over 2}\rangle_V
=\int\{d^{3}p_{2}\}\hat C^{-1/2}\phi^{\prime}(x^{\prime},k_{\perp}^{\prime})\phi(x,k_{\perp})\nonumber\\
&&
\times\sum_{\lambda_{2}}
{\rm Tr} \{u(\bar P',S'_{z})\bar{u}(\bar P',S'_{z})
\left[\bar{\Gamma}^{\prime}_{S(A)}(\strich p_{1}^{\prime}+m_{1}^{\prime})
\gamma^{\mu}(\strich p_{1}+m_{1})\Gamma_{S(A)}\right]u(\bar P,S_{z})
\bar{u}(\bar P,S_{z})(\hat{\Gamma}^{\mu}_{1\over 2})_i\}\,,
\nonumber\\
&&
(J_{5})_j\cdot \langle T_{3\over 2}\rangle_V
={\rm Tr}\{u_{\beta}(P',S'_{z})\bar{u}_{\alpha}(P',S'_{z})\nonumber\\
&&
\times\left[\frac{P^{\alpha}}{M}\left(\gamma^{\mu}F^V_{1}
+\frac{P^{\mu}}{M} F^V_{2}
+\frac{P^{\,\prime\mu}}{M^{\prime}}F^V_{3}\right)+g^{\alpha\mu}F^V_{4}\right]
\gamma_{5}\,u(P,S_{z})\bar u (P,S_{z})({\Gamma}^{\mu \beta}_{3\over 2})_j\gamma_5\}\,,
\nonumber\\
&&
(\hat J_{5})_j\cdot \langle \hat T_{3\over 2}\rangle_V
=\int\{d^{3}p_{2}\}\hat C^{-1/2}\phi^{\prime}(x^{\prime},k_{\perp}^{\prime})\phi(x,k_{\perp})
\sum_{\lambda_{2}}{\rm Tr}\{u_{\beta}(\bar P',S'_{z})\bar{u}_{\alpha}(\bar P',S'_{z})\nonumber\\
&&
\times\left[\bar{\Gamma}^{\,\prime\alpha}_{A}(\strich p_{1}^{\prime}+m_{1}^{\prime})
\gamma^{\mu}(\strich p_{1}+m_{1})\Gamma_{A}\right]u(\bar P,S_{z})
\bar u(\bar P,S_{z})(\hat{\Gamma}^{\mu \beta}_{3\over 2})_j\gamma_5\}\,.
\end{eqnarray}
Furthermore, we relate $(I)_i\cdot \langle T_{1/2}\rangle_V$ and $(J_5)_j\cdot \langle T_{3/2}\rangle_V$
to $(\hat I)_i\cdot \langle \hat T_{1/2}\rangle_V$ and $(\hat J_5)_j\cdot \langle \hat T_{3/2}\rangle_V$,
respectively, to extract $f_i^V$ and $F_j^V$.
Similarly, $(I_5)_i\cdot \langle T_{1/2}\rangle_A$, $(\hat I_5)_i\cdot \langle \hat T_{1/2}\rangle_A$,
$(J)_j\cdot \langle T_{3/2}\rangle_A$, and $(\hat J)_j\cdot \langle \hat T_{3/2}\rangle_A$
can be used to extract $f_i^A$ and $F_j^A$.

\section{Numerical Analysis}
In the numerical analysis, we use the Wolfenstein parameters
to present the CKM matrix elements,
given by~\cite{pdg}
\begin{eqnarray}\label{CKM}
&&
(V_{cb},V_{cd},V_{cs})=(A\lambda^2,-\lambda,1-\lambda^2/2)\,,\nonumber\\
&&
(V_{ub},V_{ud},V_{us})=(A\lambda^3(\rho-i\eta),1-\lambda^2/2,\lambda)\,,
\end{eqnarray}
where $A=0.826$, $\lambda=0.225$,
$\rho=0.163\pm 0.010$, and $\eta=0.361\pm 0.007$. The decay constants can be found
in Refs.~\cite{Hsiao:2015cda,Chen:2008sw,Becirevic:2013bsa,Lucha:2014xla}:
\begin{eqnarray}\label{dconst}
&&
(f_{\eta_c},f_D,f_{D_s})=(387\pm 7, 204.6\pm 5.0, 257.5\pm 4.6)~\text{MeV}\,,\nonumber\\
&&
(f_{J/\psi},f_{D^*},f_{D_s^*})=(418\pm 9, 252.2\pm 22.7, 305.5\pm 27.3)~\text{MeV}\,.
\end{eqnarray}
We adopt $(c^{eff}_1,\,c^{eff}_2)=(1.168,-0.365)$ from Refs.~\cite{Ali:1998eb,Hsiao:2014mua}.
In the generalized factorization, where
$N_c^{eff}$ is varied from 2, 3, to $\infty$, we obtain
\begin{eqnarray}\label{a1a2b}
a_1=(0.99,1.05,1.17)\,,\;a_2=(0.22,0.02,-0.37)\,,
\end{eqnarray}
which serve as the input parameters in the amplitudes in Eq.~(\ref{amp1}).

To calculate $f_i^{V(A)}$ and $F_j^{V(A)}$, the parameter inputs are needed,
given by~\cite{Ke:2019smy,Geng:2000if,Geng:2013yfa}
\begin{eqnarray}\label{mqbetaq}
(m_b,\beta_{b[ss]})&=&(5.00\pm 0.20,0.78\pm 0.04)~\mathrm{GeV}\,,\nonumber\\
(m_s,\beta_{s[ss]})&=&(0.38\pm 0.02,0.44\pm 0.02)~\mathrm{GeV}\,,\nonumber\\
(m_n,\beta_{n[ss]})&=&(0.275\pm 0.025,0.44\pm 0.02)~\mathrm{GeV}\,,
\end{eqnarray}
with $n=u$ or $d$. In Eq.~(\ref{F5j}), the connections 
$(I)_i\cdot \langle T_{1/2}\rangle_V=(\hat I)_i\cdot \langle \hat T_{1/2}\rangle_V$ and
$(J_5)_j\cdot \langle T_{3/2}\rangle_V=(\hat J_5)_j\cdot \langle \hat T_{3/2}\rangle_V$
enable us to extract the form factors in the light-front quark model~\cite{Hsiao:2020gtc,
Zhao:2018zcb,Ke:2012wa,Ke:2017eqo,Zhao:2018mrg,Hu:2020mxk,Ke:2019smy,Hsiao:2021mlp}.
Similarly, we obtain $f_i^A$ and $F_j^A$ with
$(I_5)_i\cdot \langle T_{1/2}\rangle_A=(\hat I_5)_i\cdot \langle \hat T_{1/2}\rangle_A$ and
$(J)_j\cdot \langle T_{3/2}\rangle_A=(\hat J)_j\cdot \langle \hat T_{3/2}\rangle_A$, 
respectively. 

The light-front quark model yields momentum-dependent form factors, which can be reproduced 
using the dipole expression~\cite{Zhao:2018zcb,Hsiao:2020gtc,Hsiao:2021mlp,Hsiao:2019wyd}
\begin{eqnarray}\label{LFparameters}
F(p^2)=\frac{F_0}{1+a\left(p^2/m_{V(A)}^2\right)+b\left(p^4/m_{V(A)}^4\right)}\,,\;
\end{eqnarray}
where $(a,b)$ are fit parameters, $F_0\equiv F(0)$ represents the form factor at $p^2=0$, 
and the pole mass $m_{V(A)}$ corresponds to the (axial-)vector current of the $b\to q$ transition.

Since and the ground state $\Omega_b^-$ and
the excited state $\Omega_b(6316)^-$~\cite{LHCb:2020tqd}
are assigned with parity-even and parity-odd quantum numbers, 
respectively~\cite{pdg,Karliner:2020fqe,Ali:2020tvs}, in accordance 
with the vector and axial-vector currents~\cite{Cheng:1996cs,Becirevic:1999kt,Hsiao:2023qtk},
we adopt their masses as $(m_V, m_A)=(6.05,6.32)$~GeV, respectively.
%
\begin{figure}[t!]
\includegraphics[width=3in]{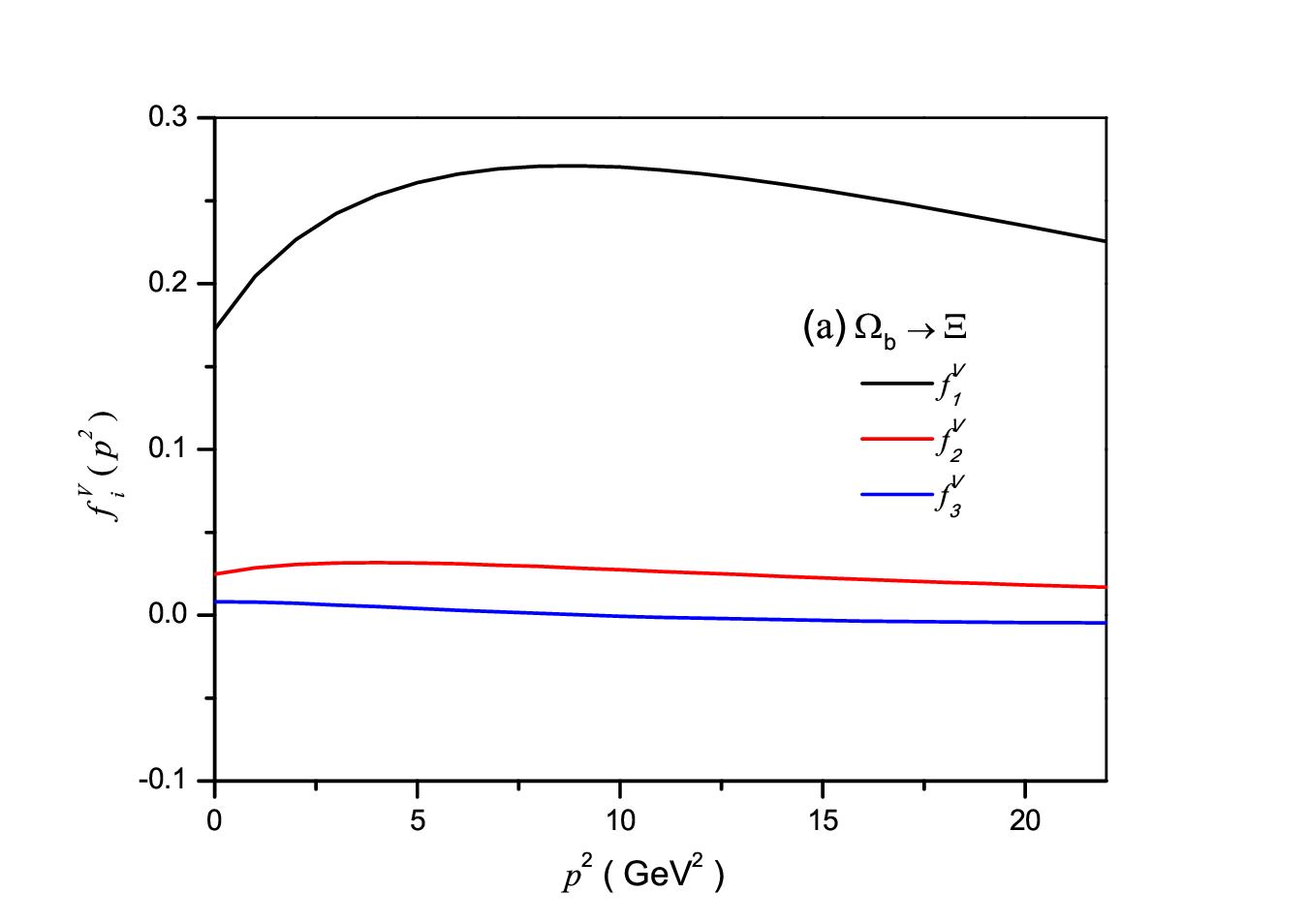}
\includegraphics[width=3in]{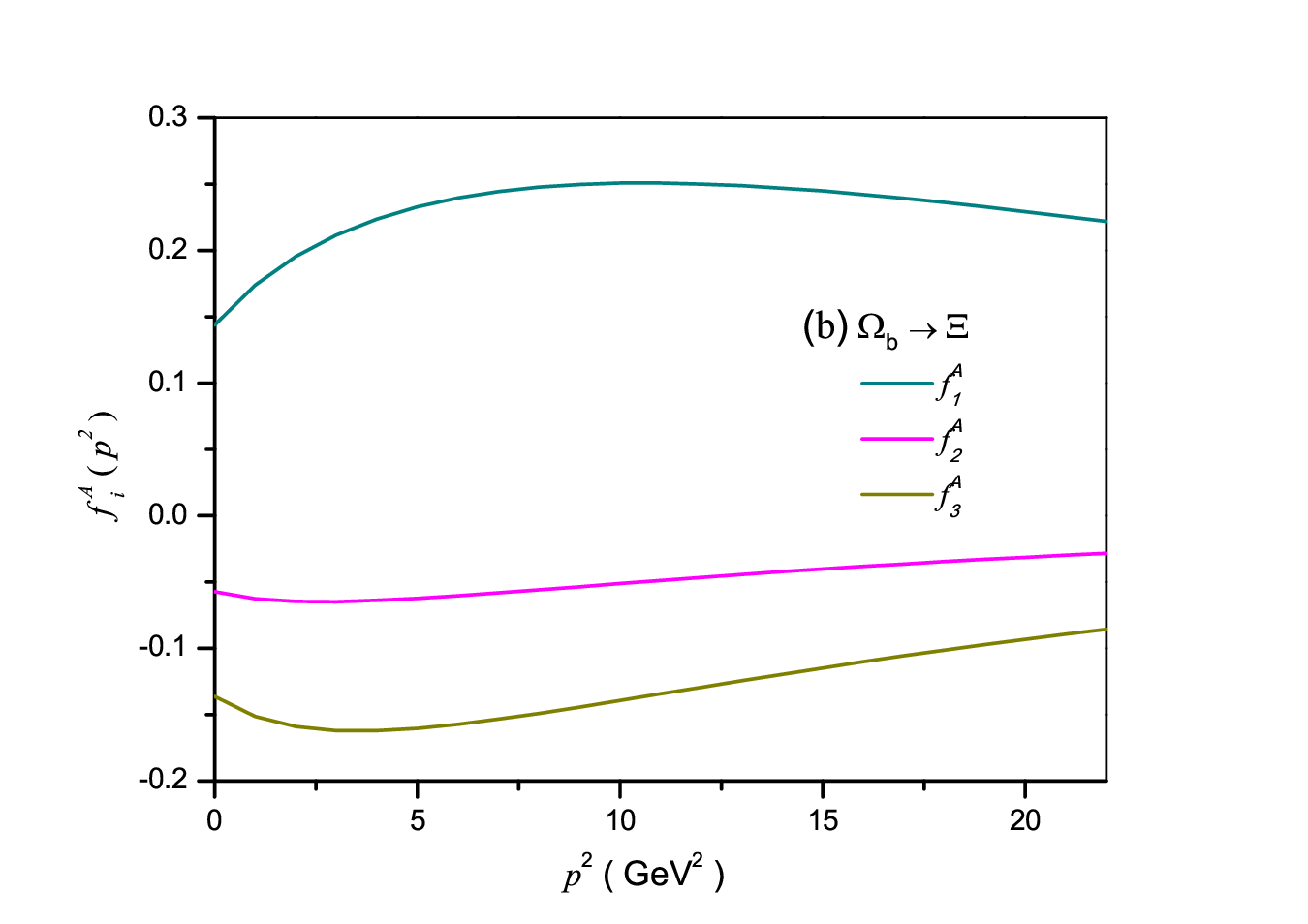}
\includegraphics[width=3in]{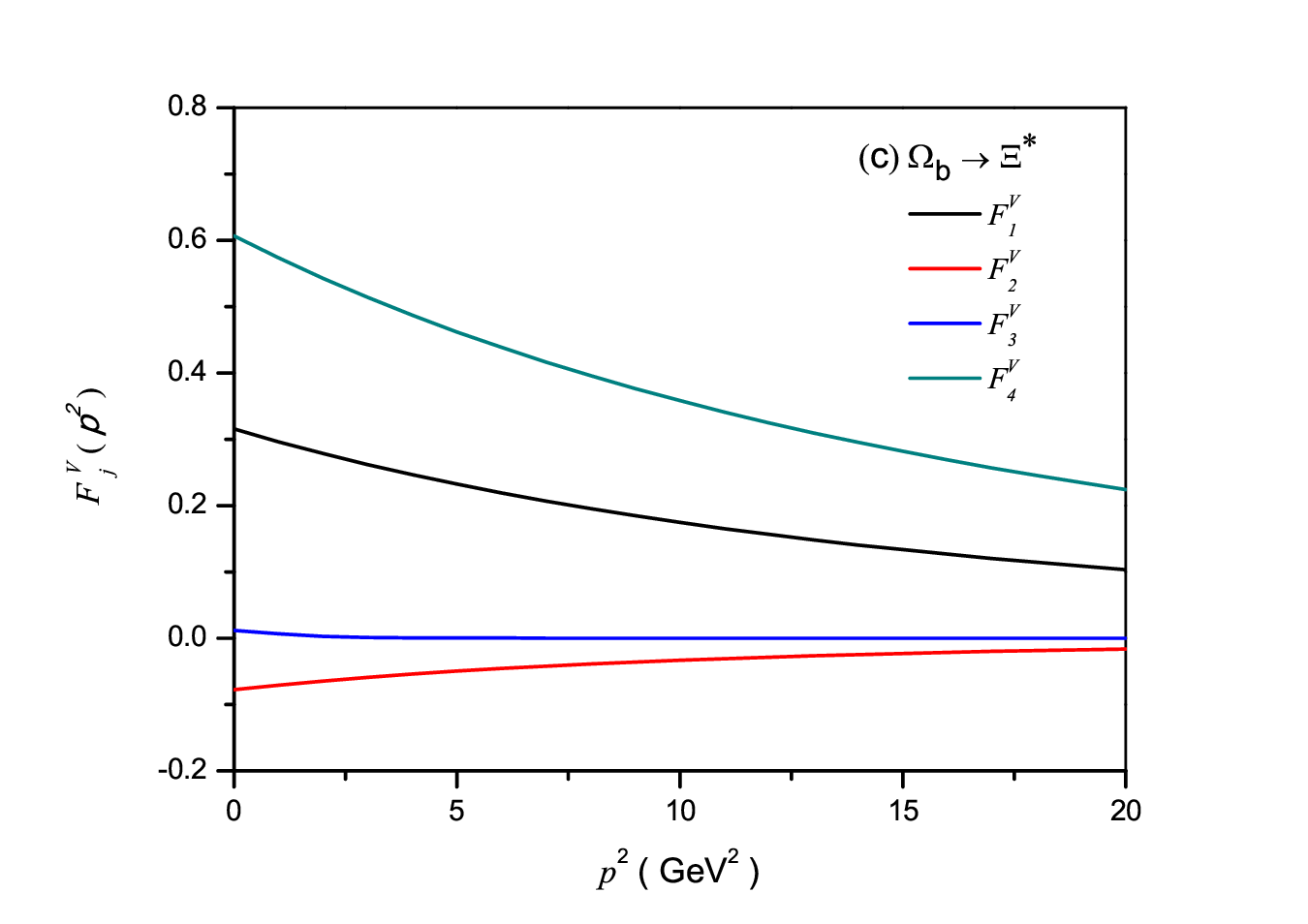}
\includegraphics[width=3in]{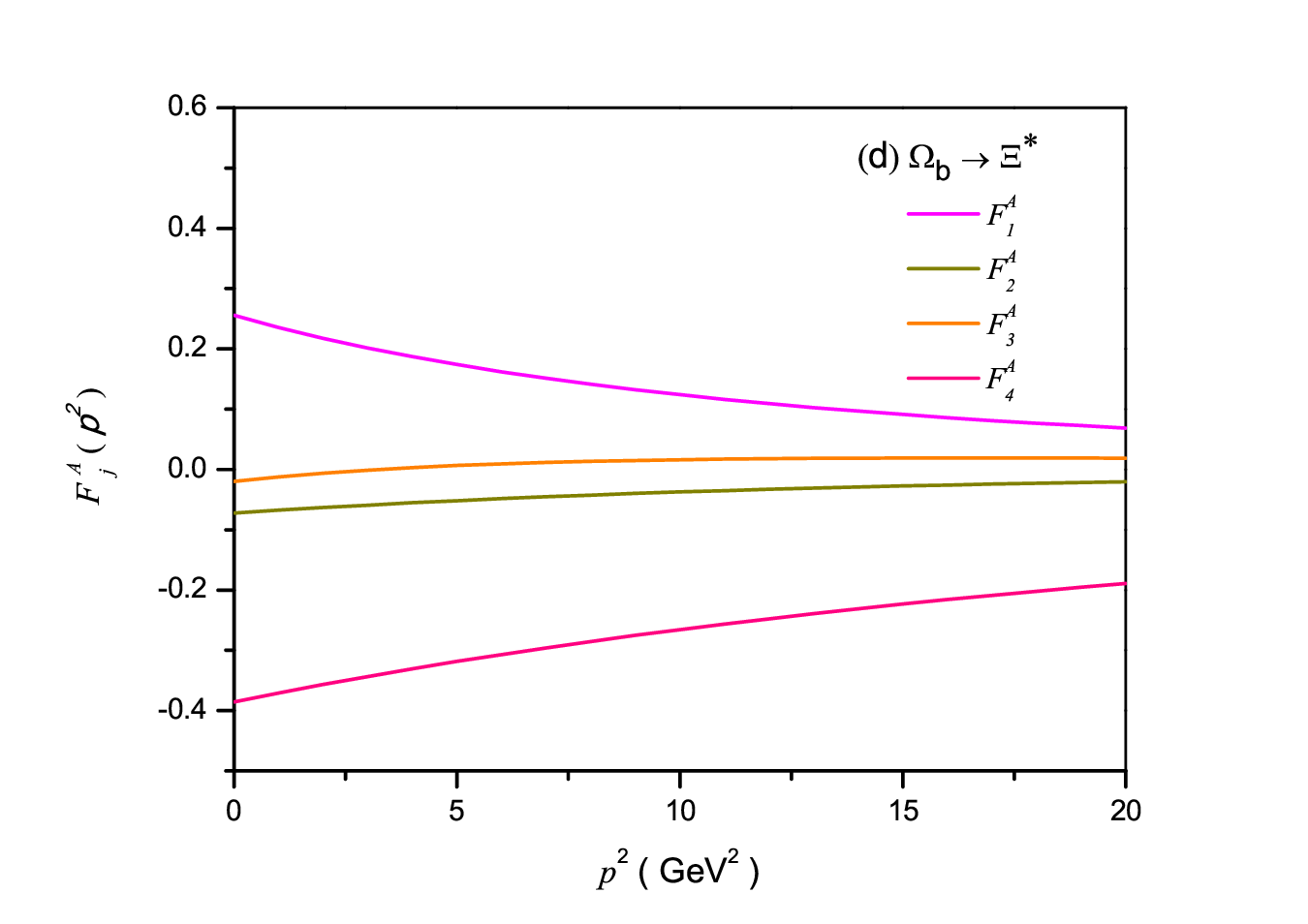}
\includegraphics[width=3in]{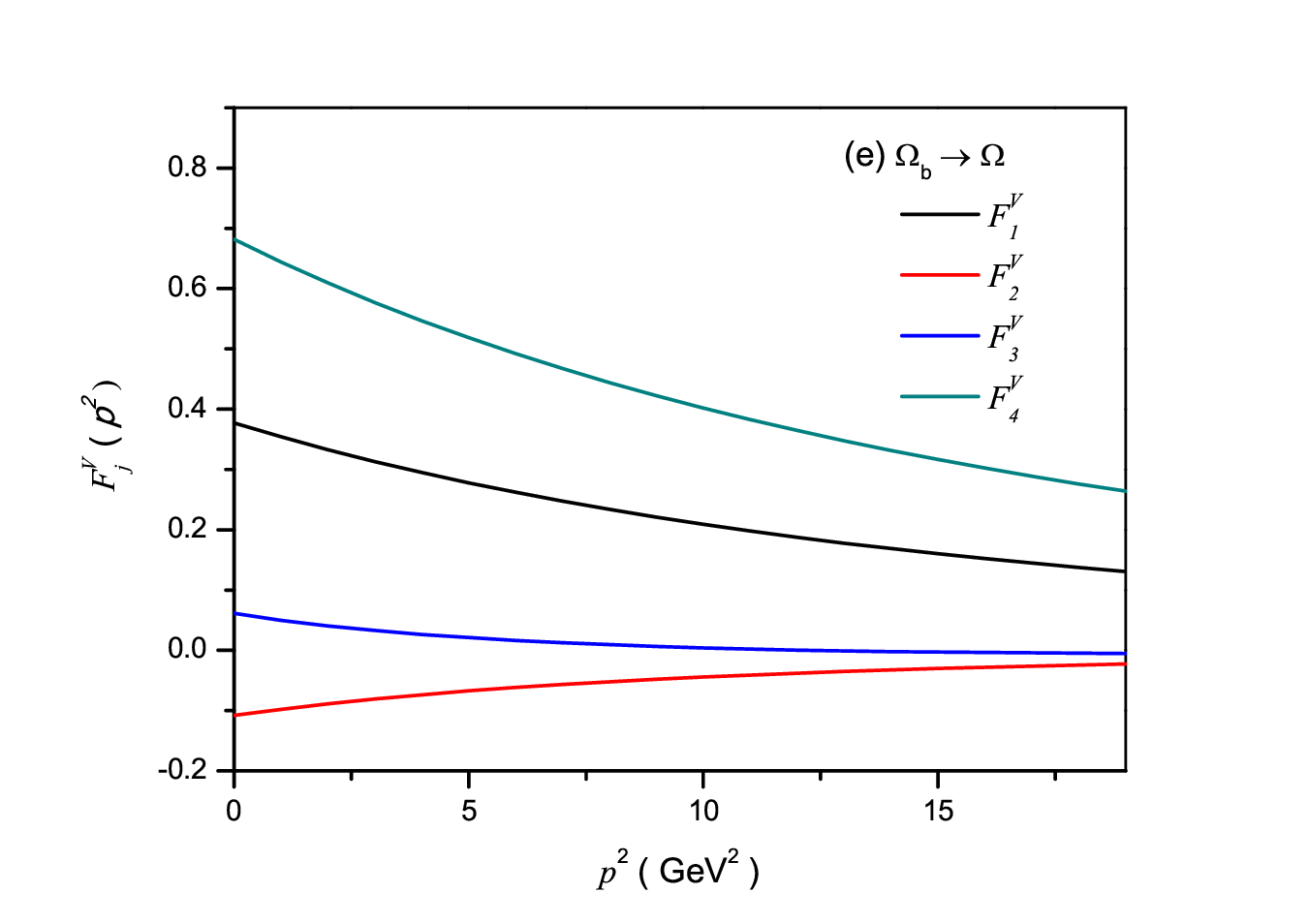}
\includegraphics[width=3in]{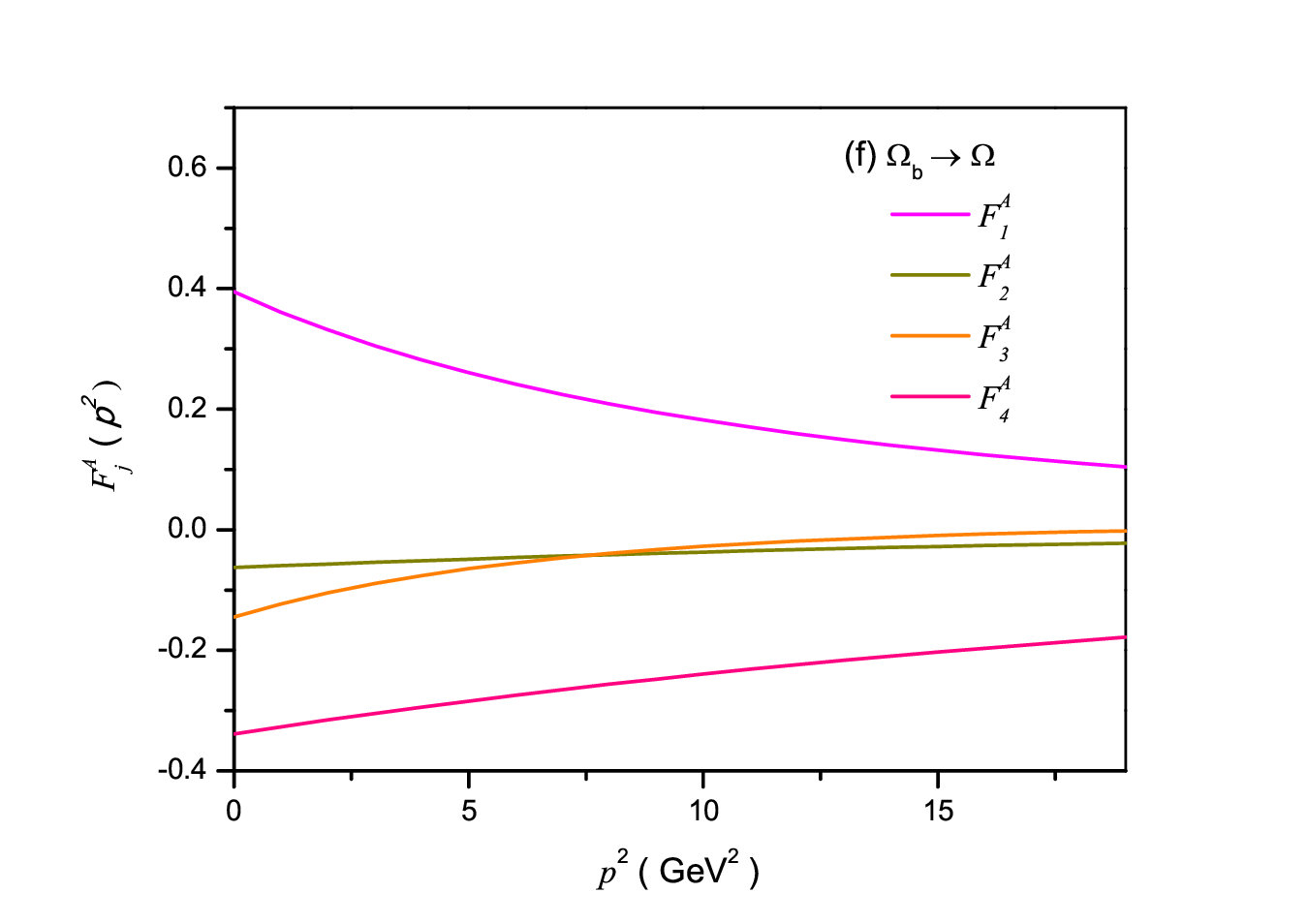}
\caption{The $p^2$-dependent form factors, calculated in the light-front quark model, 
are depicted as follows:
(a)~$f^V_i(p^2)$ and (b)~$f^A_i(p^2)$ for the $\Omega_b^-\to \Xi$ transition,
(c)~$F^V_j(p^2)$ and (d)~$F^A_j(p^2)$ for the $\Omega_b^-\to \Xi^*$ transition, and
(e)~$F^V_j(p^2)$ and (f)~$F^A_j(p^2)$ for the $\Omega_b^-\to \Omega^-$ transition.}\label{fig2}
\end{figure}
%
%
\begin{table}[b!]
\caption{The form factors $f^{V,A}_i$ for the $\Omega_b\to\Xi$ transition,
with $(F_0,a,b)$ calculated in the light-front quark model.
The errors incorporate the uncertainties from $m_{q_1}$ and $\beta_{q_1[ss]}$
in Eq.~(\ref{mqbetaq}), the same as those in Tables~\ref{tab1b} and \ref{tab2}. Additionally,
we present $f_i^{V,A}$ at $p^2=(m_{\eta_c(J/\psi)}^2,m_{D^{(*)}}^2)$.}\label{tab1}
{
\tiny
\begin{tabular}{|c|c|c|c|} \hline
& $(F_0,a,b)$ 
& $p^2=(m_{\eta_c}^2,m_{J/\psi}^2)$ 
& $p^2=(m_{D}^2,m_{D^*}^2)$
    \\ \hline \hline
$f^V_{1}$ 
& $(0.172\pm 0.015,-2.53\pm0.21,4.13\pm0.53)$ 
& $(0.274\pm0.036 ,0.277\pm0.040)$ 
& $(0.216\pm0.020,0.223\pm0.021)$
\\
$f^V_{2}$ 
& $(0.024\pm 0.001,-2.06\pm0.19,5.62\pm0.36)$ 
& $(0.029\pm0.002 ,0.028\pm0.002)$ 
& $(0.028\pm0.001, 0.029\pm0.002)$
\\
$f^V_{3}$ 
& $(0.008\pm 0.004,-4.37\pm0.41,104.72\pm19.72)$ 
& $\sim (0,0)$ 
& $\sim (0,0)$
\\\hline
$f^A_{1}$ 
& $(0.142\pm 0.020,-2.83\pm0.08,4.37\pm0.19)$ 
& $(0.242\pm0.035,0.248\pm0.043)$ 
& $(0.181\pm0.026 ,0.187^{+0.032}_{-0.022})$
\\ 
$f^A_{2}$ 
& $(-0.058\pm 0.015,-1.22\pm0.01,6.25\pm0.50)$ 
& $(-0.056\pm0.015,-0.054\pm0.015)$ 
& $(-0.062\pm0.016,-0.062\pm0.016)$
\\
$f^A_{3}$ 
& $(-0.136\pm 0.030,-1.59\pm0.02,5.45\pm0.40)$ 
& $(-0.148\pm0.033,-0.145^{+0.034}_{-0.028})$ 
& $(-0.151\pm0.034,-0.152\pm0.034)$
\\\hline
\end{tabular}}
\end{table}
%
%
\begin{table}[b!]
\caption{The form factors $F_j^{V,A}$ for the $\Omega_b\to\Xi^{*}$ transition,
presented with $(F_0,a,b)$ and
at $p^2=(m_{\eta_c(J/\psi)}^2,m_{D^{(*)}}^2)$.}\label{tab1b}
{
\tiny
\begin{tabular}{|c|c|c|c|} \hline
& $(F_0,a,b)$ 
& $p^2=(m_{\eta_c}^2,m_{J/\psi}^2)$ 
& $p^2=(m_{D}^2,m_{D^*}^2)$
    \\ \hline \hline
$F^V_{1}$ 
& $(0.315\pm 0.002,2.24\pm0.05,2.63\pm0.32)$ 
& $(0.185\pm0.002,0.178\pm0.002)$ 
& $(0.255\pm0.002,0.246\pm0.002)$
\\ 
$F^V_{2}$ 
& $(-0.077\pm 0.001,3.29\pm0.11,6.17\pm0.87)$ 
& $(-0.036\pm0.001,-0.034\pm0.001)$ 
& $(-0.056\pm0.002,-0.054\pm0.002)$
\\ 
$F^V_{3}$ 
& $(0.012\pm 0.004,-7.56\pm6.47,1274.08\pm78.53)$ 
& $\sim (0,0)$ 
& $\sim (0,0)$
\\ 
$F^V_{4}$ 
& $(0.607\pm 0.008,2.01\pm0.01,1.94\pm0.22)$ 
& $(0.378\pm0.006 ,0.365\pm0.006)$ 
& $(0.502\pm0.006 ,0.488\pm0.007)$
\\ \hline
$F^A_{1}$ 
& $(0.255\pm 0.005,3.21\pm0.03,4.23\pm0.08)$ 
& $(0.132\pm0.003 ,0.126\pm0.003)$ 
& $(0.194\pm0.004, 0.187\pm0.004)$
\\ 
$F^A_{2}$ 
& $(-0.072\pm 0.002,2.54\pm0.14,4.86\pm0.11)$ 
& $(-0.040\pm0.001,-0.038\pm0.001)$
&$(-0.057\pm0.002,-0.055\pm0.002)$
\\ 
$F^A_{3}$ 
& $(-0.020\pm 0.007,-10.53\pm7.62,1349.04\pm871.00)$ 
& $\sim (0,0)$ 
& $\sim (0,0)$
\\ 
$F^A_{4}$ 
& $(-0.385\pm 0.011,1.55\pm0.06,1.02\pm0.03)$ 
& $(-0.276\pm0.008,-0.269\pm0.009)$ 
& $(-0.337\pm0.010,-0.330\pm0.009)$
\\\hline
\end{tabular}}
\end{table}
%
%
\begin{table}[b!]
\caption{The form factors $F_j^{V,A}$ for the $\Omega_b\to\Omega$ transition.}\label{tab2}
{
\tiny
\begin{tabular}{|c|c|c|c|} \hline
    & $(F_0,a,b)$ & $p^2=(m_{\eta_c}^2,m_{J/\psi}^2)$ & $p^2=(m_{D}^2,m_{D^*}^2)$
\\ \hline \hline
$F^V_{1}$ & $(0.377\pm 0.015,2.25\pm0.06,2.56\pm0.33)$ & $(0.222\pm0.009,0.213\pm0.010)$ & $ (0.305\pm0.012,0.295\pm0.011)$
\\ 
$F^V_{2}$ & $(-0.108\pm 0.004,3.48\pm0.21,6.73\pm0.87)$ & $(-0.048\pm0.002,-0.045\pm0.003)$ & $(-0.078\pm0.004,-0.074\pm0.005)$
\\ 
$F^V_{3}$ & $(0.061\pm 0.004,3.41\pm0.51,95.12\pm22.66)$ & $(0.008\pm0.002,0.007\pm0.002)$ & $(0.028\pm0.004,0.024\pm0.005)$
\\ 
$F^V_{4}$ & $(0.682\pm 0.035,2.03\pm0.03,1.89\pm0.26)$ & $(0.424\pm0.023,0.410\pm0.022)$ & $(0.563\pm0.029,0.547\pm0.028)$
\\ \hline
$F^A_{1}$ & $(0.394\pm 0.027,3.50\pm0.02,4.76\pm0.14)$ & $(0.195\pm0.014,0.186\pm0.013)$ & $(0.294\pm0.020,0.281\pm0.019)$
\\ 
$F^A_{2}$ & $(-0.062\pm 0.01,1.68\pm0.18,4.27\pm0.02)$ & $(-0.039\pm0.006,-0.038\pm0.006)$ & $(-0.053\pm0.009,-0.051\pm0.009)$
\\ 
$F^A_{3}$ & $(-0.145\pm 0.028,3.85\pm0.88,56.15\pm32.12)$ & $(-0.031^{+0.010}_{-0.017},-0.028^{+0.009}_{-0.017})$ & $(-0.082\pm0.022,-0.074\pm0.021)$
\\ 
$F^A_{4}$ & $(-0.339\pm 0.046,1.40\pm0.05,1.01\pm0.04)$ & $(-0.249\pm0.034,-0.243\pm0.033)$ & $(-0.300\pm0.041,-0.294\pm0.040)$
\\\hline
\end{tabular}}
\end{table}
%
%
\begin{table}[b]
\caption{Branching fractions of two-body charmful $\Omega_b^-$ decays
with $N_c^{eff}=(2,3,\infty)$, which estimates the non-factorizable QCD corrections,
are presented in comparison with other calculations.  
The error bars in our results include the uncertainties from the CKM matrix elements 
in Eq.~(\ref{CKM}), the decay constants in Eq.~(\ref{dconst}), and the form factors
evaluated at $p^2=(m^2_{\eta_c},m^2_{J/\psi},m^2_{D^{(*)}})$, 
as given in Tables~\ref{tab1}, \ref{tab1b}, and \ref{tab2}.}\label{tab3}
\begin{center}
{
\tiny
\begin{tabular}{|l|c|c|c|}
\hline
Branching fraction
&$b$-decays
&This work: $N_c^{eff}=(2, 3, \infty)$
&Gutsche~\cite{Gutsche:2018utw}\\
\hline\hline
${\cal B}(\Omega_b^-\to\Xi^- D^0)$
&$b\to c\bar u d$
&$(1.4\pm0.3,0.012\pm0.003,4.1\pm0.9)\times10^{-4}$
&$1.4\times 10^{-6}$\\
${\cal B}(\Omega_b^-\to\Xi^{-}D^{*0})$
&$b\to c\bar u d$
&$(3.0\pm0.9,0.02\pm0.01,8.4^{+2.5}_{-2.0})\times10^{-4}$
&$2.1\times 10^{-6}$\\
${\cal B}(\Omega_b^-\to\Xi^0 D_s^-)$
&$b\to u \bar c s$
&$(4.0\pm0.9, 4.5\pm1.0,5.5\pm1.2)\times10^{-5}$
&---\\
${\cal B}(\Omega_b^-\to\Xi^{0}D_s^{*-})$
&$b\to u \bar c s$
&$(7.8^{+2.4}_{-1.9}, 8.7^{+2.7}_{-2.1},10.9^{+3.3}_{-2.6})\times10^{-5}$
&---\\
${\cal B}(\Omega_b^-\to\Xi^-\eta_c)$
&$b\to c\bar c d$
&$(3.1\pm0.8,0.026\pm0.007,8.9^{+2.3}_{-2.0})\times10^{-5}$
&$0.13\times10^{-6}$\\
${\cal B}(\Omega_b^-\to\Xi^{-} J/\psi)$
&$b\to c\bar c d$
&$(6.0^{+1.7}_{-1.5},0.05\pm 0.01,17.1^{+4.8}_{-4.2})\times10^{-5}$
&$0.18\times10^{-6}$\\
${\cal B}(\Omega_b^-\to\Xi^0 D^-)$
&$b\to u \bar c d$
&$(1.3\pm 0.3, 1.5\pm0.3,1.8\pm0.4)\times10^{-6}$
&$1.6\times 10^{-8}$\\
${\cal B}(\Omega_b^-\to\Xi^{0}D^{*-})$
&$b\to u \bar c d$
&$(2.7\pm 0.8, 3.0\pm 0.2\pm0.9,3.8^{+1.2}_{-0.9})\times10^{-6}$
&$2.6\times 10^{-8}$\\
${\cal B}(\Omega_b^-\to\Xi^- \bar D^0)$
&$b\to u \bar c d$
&$(6.4\pm0.4^{+1.5}_{-1.3},0.05\pm0.01,18.2^{+4.2}_{-3.8})\times10^{-8}$
&$7.3\times 10^{-10}$\\
${\cal B}(\Omega_b^-\to\Xi^{-}\bar D^{*0})$
&$b\to u \bar c d$
&$(13.3^{+4.1}_{-3.2},0.11\pm0.04,37.5^{+11.5}_{-\;\,9.2})\times10^{-8}$
&$13.2\times 10^{-10}$\\
\hline\hline
${\cal B}(\Omega_b^-\to\Xi^{*-} D^{0})$
&$b\to c\bar u d$
&$(3.1\pm0.4,0.03\pm0.01,8.7\pm1.0)\times10^{-4}$
&$0.7\times 10^{-7}$\\
${\cal B}(\Omega_b^-\to\Xi^{*-} D^{*0})$
&$b\to c\bar u d$
&$(6.6\pm1.3,0.05\pm0.01,18.6^{+3.7}_{-3.4})\times10^{-4}$
&$1.9\times 10^{-7}$\\
${\cal B}(\Omega_b^-\to\Xi^{*0} D_s^{-})$
&$b\to u \bar c s$
&$(7.7\pm1.0,8.7\pm1.1,10.8\pm1.3)\times10^{-5}$
&$0.49\times10^{-6}$\\
${\cal B}(\Omega_b^-\to\Xi^{*0} D_s^{*-})$
&$b\to u \bar c s$
&$(15.8^{+3.3}_{-3.0},17.8^{+3.7}_{-3.4},22.1^{+4.6}_{-4.2})\times10^{-5}$
&$1.13\times10^{-6}$\\
${\cal B}(\Omega_b^-\to\Xi^{*-}\eta_c)$
&$b\to c\bar c d$
&$(2.0\pm0.2,0.016\pm0.002,5.6\pm0.6)\times10^{-5}$
&$0.69\times10^{-5}$\\
${\cal B}(\Omega_b^-\to\Xi^{*-} J/\psi)$
&$b\to c\bar c d$
&$(4.7\pm0.4,0.039\pm0.004,13.3\pm1.1)\times10^{-5}$
&$2.0\times10^{-5}$\\
${\cal B}(\Omega_b^-\to\Xi^{*0} D^{-})$
&$b\to u \bar c d$
&$(2.8\pm 0.4, 3.2\pm 0.4,3.9\pm0.5)\times10^{-6}$
&$6.7\times 10^{-8}$\\
${\cal B}(\Omega_b^-\to\Xi^{*0} D^{*-})$
&$b\to u \bar c d$
&$(6.0^{+1.5}_{-1.1}, 6.8^{+1.4}_{-1.3},8.4\pm1.7)\times10^{-6}$
&$18.5\times 10^{-8}$\\
${\cal B}(\Omega_b^-\to\Xi^{*-} \bar D^{0})$
&$b\to u \bar c d$
&$(13.7\pm1.8,0.11\pm0.02,38.9^{+5.0}_{-4.7})\times10^{-8}$
&$6.8\times10^{-8}$\\
${\cal B}(\Omega_b^-\to\Xi^{*-} \bar D^{*0})$
&$b\to u \bar c d$
&$(29.6^{+6.1}_{-5.6},0.24^{+0.06}_{-0.04},83.6^{+17.3}_{-15.9})\times10^{-8}$
&$18.9\times10^{-8}$\\
\hline\hline
${\cal B}(\Omega_b^-\to\Omega^- \eta_c)$
&$b\to c\bar c s$
&$(2.1^{+1.8}_{-1.0},0.02\pm0.01,5.9^{+5.0}_{-3.0})\times10^{-4}$
&$1.9\times10^{-4}$\\
${\cal B}(\Omega_b^-\to\Omega^- J/\psi)$
&$b\to c\bar c s$
&$(6.9^{+2.4}_{-1.7},0.06\pm0.02,19.6^{+6.9}_{-4.7})\times10^{-4}$
&$8.1\times10^{-4}$\\
${\cal B}(\Omega_b^-\to\Omega^- D^{0})$
&$b\to c\bar u s$
&$(1.6^{+1.2}_{-0.8},0.01\pm0.01,4.4^{+3.5}_{-2.3})\times10^{-5}$
&$0.07\times10^{-6}$\\
${\cal B}(\Omega_b^-\to\Omega^- D^{*0})$
&$b\to c\bar u s$
&$(3.4^{+1.8}_{-1.3},0.03\pm0.01,9.6^{+5.0}_{-3.6})\times10^{-5}$
&$0.20\times10^{-6}$\\
${\cal B}(\Omega_b^-\to\Omega^- \bar D^{0})$
&$b\to u\bar c s$
&$(2.4^{+1.9}_{-1.3},0.02\pm0.02,6.9^{+5.4}_{-3.6})\times10^{-6}$
&$0.1\times10^{-6}$\\
${\cal B}(\Omega_b^-\to\Omega^- \bar D^{*0})$
&$b\to u\bar c s$
&$(5.3^{+2.8}_{-2.0},0.04\pm0.03,15.0^{+8.0}_{-5.7})\times10^{-6}$
&$0.3\times10^{-6}$\\
\hline
\end{tabular}}
\end{center}
\end{table}
%

\newpage
Utilizing the connections:
$(I)_i\cdot \langle T_{1/2}\rangle_V=(\hat I)_i\cdot \langle \hat T_{1/2}\rangle_V$ and 
$(J_5)_j\cdot \langle T_{3/2}\rangle_V=(\hat J_5)_j\cdot \langle \hat T_{3/2}\rangle_V$,
we extract $f_i^V$ and $F_j^V$, respectively. Likewise, we extract 
$f_i^A$ from $(I_5)_i\cdot \langle T_{1/2}\rangle_A=(\hat I_5)_i\cdot \langle \hat T_{1/2}\rangle_A$
and $F_j^A$ from $(J)_j\cdot \langle T_{3/2}\rangle_A=(\hat J)_j\cdot \langle \hat T_{3/2}\rangle_A$.
The extracted results, as shown in Fig.~\ref{fig2}, exhibit a clear dependence on $p^2$.
As listed in Tables~\ref{tab1}, ~\ref{tab1b}, and~\ref{tab2}, 
we further determine the parameters $(F_0, a, b)$ in the dipole expression of Eq.~(\ref{LFparameters})
by fitting to the data points along the curves shown in Fig.~\ref{fig2}.  Then we provide
the form factors at $p^2=(m_{\eta_c}^2,m_{J/\psi}^2,m^2_{D^{(*)}})$, finding that
$F(m^2_{D^{(*)}})\simeq F(m^2_{D_s^{(*)}})$. Subsequently, the branching fractions of 
the singly and doubly charmful two-body $\Omega_b^-$ decays
are calculated, as summarized in Table~\ref{tab3}.

\section{Discussions and Conclusion}
The $\Omega_b^-\to \Xi^{(*)}$ and $\Omega_b^-\to\Omega^-$ transition form factors,
denoted by $F_{\Xi^{(*)}}$ and $F_{\Omega}$, respectively, have been partially studied 
in various models~\cite{Zhao:2018zcb,Hsiao:2021mlp,Cheng:1996cs}. In this work,
we propose using the light-front quark model to calculate
$F_{\Xi}$, $F_{\Xi^*}$ and $F_{\Omega}$, which are applied to
investigate two-body singly and doubly charmful $\Omega_b^-$ decays.
As a result, the predicted branching fractions span a range from
$10^{-4}$ to $10^{-8}$, as listed in Table~\ref{tab3}. These values are consistent with 
the expected suppression due to the CKM matrix elements. For example, 
$|V_{cb}V_{cs}^*|$ and $|V_{cb}V_{ud}^*|$ correspond to ${\cal B}\simeq 10^{-4}$, 
while the more suppressed $|V_{ub}V_{cd}^*|$ leads to a much smaller ${\cal B}\simeq 10^{-8}$.

It appears that the branching fractions
${\cal B}(\Omega_b^-\to\Xi^{*-}\eta_c,\Xi^{*-} J/\psi)$,
${\cal B}(\Omega_b^-\to\Xi^{*-} \bar D^{(*)0})$, and
${\cal B}(\Omega_b^-\to\Omega^- \eta_c,\Omega^- J/\psi)$ are consistent with
the results of the previous study~\cite{Gutsche:2018utw}.
However, most of branching fractions in our work exhibit significant deviations.
For instance, ${\cal B}(\Omega_b^-\to\Xi^- D^{(*)0})$ is found to be 100 times larger than 
the corresponding value reported in Ref.~\cite{Gutsche:2018utw}.
Nevertheless, Ref.~\cite{Gutsche:2018utw} does not 
provide details on the form factors, which limits our ability to identify the source of the discrepancy.
On the other hand, the available form factors from Refs.~\cite{Zhao:2018zcb,Cheng:1996cs}
can be used for a more direct comparison between our results and those of Ref.~\cite{Gutsche:2018utw}. 

Specifically, the dominant $\Omega_b^-\to\Xi$ transition form factors
at $p^2=m_D^2$ are given as: i)~$(f_1^V,f_1^A)=(0.13,-0.04)$ and
ii)~$(f_1^V,f_1^A)=(0.23,-0.04)$, from Refs.~\cite{Cheng:1996cs}~and~\cite{Zhao:2018zcb},
respectively, where in both cases, the value of $f_1^A$ is relatively small. 
Using $a_2=0.22$ in Eq.~(\ref{a1a2b}), along with the form factors from cases i) and ii), 
we obtain ${\cal B}(\Omega_b^-\to\Xi^- D^0)=0.3 \times10^{-4}$ and $0.9 \times10^{-4}$, 
respectively. These values are approximately 20 and 60  times larger than 
the value given in~\cite{Gutsche:2018utw}.

In our work, the dominant $\Omega_b^-\to\Xi$ form factors at $p^2=m_D^2$ 
are $(f_1^V,f_1^A)=(0.216,0.181)$.
Notably, $f_1^A$ is comparable to $f_1^V$, resulting in an enhanced
branching fraction of ${\cal B}(\Omega_b^-\to\Xi^- D^{0})=1.4\times10^{-4}$,
which is larger than the corresponding values from cases i) and ii).
We therefore demonstrate that our estimation of the branching fractions is reasonable,
despite being $10-100$ times greater than those reported in Ref.~\cite{Gutsche:2018utw}.
Furthermore, the predicted branching fractions in Refs.~\cite{Zhao:2018zcb,Cheng:1996cs}
are found to be consistently larger than those given in Ref.~\cite{Gutsche:2018utw}.

According to Table~\ref{tab3}, there exist certain ratios between branching fractions,
such as ${\cal B}(\Omega_b^-\to \Xi^- D^{*0})/{\cal B}(\Omega_b^-\to \Xi^- D^{0})\simeq 2$.
For our understanding, the helicity amplitudes in Eqs.~(\ref{amp2a})~and~(\ref{amp2b})
can be useful. We thus decompose ${\cal B}(\Omega_b^-\to \Xi^- D^{0},\Xi^- D^{*0})$
as
\begin{eqnarray}\label{H0}
|H_{\Xi D}|^2&\simeq& 2C_\Xi[(1-C^{+}_{D})(f_1^{V})^2+(1-C^{-}_{D})(f_1^{A})^2]\,,\nonumber\\
|H_{\Xi D^*}|^2&\simeq&
2C_\Xi\{[(1-C^{-}_{D^*})(f_1^{V})^2+(1-C^{+}_{D^*})(f_1^{A})^2]\nonumber\\
&&+2[C_{D^*}^+(1-C^{-}_{D^*})(f_1^{V})^2+C_{D^*}^-(1-C^{+}_{D^*})(f_1^{A})^2]\}\,,
\end{eqnarray}
with $C_\Xi=(m_{\Omega_b}+m_{\Xi})^2(m_{\Omega_b}-m_{\Xi})^2/m_{\Xi}^2$
and $C^{\pm}_{M}=m_{M}^2/(m_{\Omega_b}\pm m_{\Xi})^2$. In the above equation,
$|H_{\Xi D}|^2$ results from $2(|H^V_{\bar 0}|^2+|H^A_{\bar 0}|^2)$, and
the first(second) term in $|H_{\Xi D^*}|^2$ is due to $2(|H^V_{0(1)}|^2+|H^A_{0(1)}|^2)$.
The nearly equalities of $f_1^V\simeq f_1^A$ and $C^\pm_{D^*}\simeq C^\pm_{D}$ lead to
$(|H^V_{0}|^2+|H^A_{0}|^2)/(|H^V_{\bar 0}|^2+|H^A_{\bar 0}|^2)\simeq 0.98$.
Additionally, the suppression of the pre-factors $(C_{D^*}^+,C_{D^*}^-)=(0.07,0.18)$
makes $(|H^V_1|^2+|H^A_1|^2)/(|H^V_0|^2+|H^A_0|^2)\simeq 0.26$.
We hence obtain
\begin{eqnarray}\label{R0}
{\cal R}_1\equiv\frac{{\cal B}(\Omega_b^-\to\Xi^{-} D^{*0})}{{\cal B}(\Omega_b^-\to\Xi^{-}D^0)}
\simeq\frac{(m_{D^*} f_{D^*})^2 |H_{\Xi D^*}|^2}{(m_D f_{D})^2 |H_{\Xi D}|^2}\simeq 2\,,
\end{eqnarray}
with $(m_{D^*} f_{D^*})^2/(m_D f_D)^2\simeq 1.76$ and
$|H_{\Xi D^*}|^2/|H_{\Xi D}|^2$ $\simeq 1.24$, 
which agrees with the calculation in Table~\ref{tab3}. On the other hand,
${\cal R}_1$ of Ref.~\cite{Gutsche:2018utw} being 1.5 indicates that
$|H_{\Xi D^*}|^2$ must be smaller than $|H_{\Xi D}|^2$,
which seems unachievable.

For ${\cal B}(\Omega_b^-\to \Xi^{*-} D^0)$, utilizing $H_{B^*P}$ of Eqs.~(\ref{amp2a})~and~(\ref{amp2b})
leads to
\begin{eqnarray}\label{H1}
&&
|H_{\Xi^*D}|^2=2(|H^V_{{1\over 2}\bar 0}|^2+|H^A_{{1\over 2}\bar 0}|^2)\,,\nonumber\\
&&
|H^{V(A)}_{{1\over 2}\bar 0}|^2=C^\mp_{\Xi^*D}m_{\Omega_b}^2
\bigg[\bigg(1\pm \frac{m_{\Xi^*}}{m_{\Omega_b}}\bigg)F_1^{V(A)}
\mp D^+_{\Xi^*D} F_2^{V(A)}
\mp D^-_{\Xi^*D} F_3^{V(A)}
\mp F_4^{V(A)}\bigg]^2\,,
\end{eqnarray}
where
$C^\pm_{{\bf B}^*M}=
[(m_{\Omega_b}^2-m_{{\bf B}^*}^2)^2-2m_M^2 (m_{\Omega_b}^2+m_{{\bf B}^*}^2)+m_M^4]
[(m_{\Omega_b}\pm m_{{\bf B}^*})^2-m_M^2]/[6(m_{\Omega_b}m_{{\bf B}^*}m_{M})^2]$,
$D^+_{{\bf B}^*M}=(m_{\Omega_b}^2-m_{{\bf B}^*}^2+m_M^2)/(2 m_{\Omega_b}^2)$, and
$D^-_{{\bf B}^*M}=(m_{\Omega_b}^2-m_{{\bf B}^*}^2-m_M^2)/(2 m_{\Omega_b} m_{{\bf B}^*})$.
In Eq.~(\ref{H1}), $(C^-_{\Xi^*D},C^+_{\Xi^*D})=(8.1,26.0)$,
$(D^+_{\Xi^*D},D^-_{\Xi^*D})=(0.5,1.7)$, and $F_3^{V(A)}$ is negligible.
Since $F_1^{V}$ substantially cancels $F_4^{V}$ in $|H^{V}_{{1\over 2}\bar 0}|^2$,
we obtain $|H^V_{{1\over 2}\bar 0}|^2/|H^A_{{1\over 2}\bar 0}|^2\simeq 0.15$, such that
$|H^A_{{1\over 2}\bar 0}|^2$ becomes the dominant contribution.

For ${\cal B}(\Omega_b^-\to \Xi^{*-} D^{*0})$,
we make the decomposition
\begin{eqnarray}\label{H2}
&&
|H_{\Xi^*D^*}|^2=2(|H^V_{{1\over 2}0}|^2+|H^A_{{1\over 2}0}|)^2
+2(|H^V_{{1\over 2}1}|^2+|H^A_{{1\over 2}1}|^2)
+2(|H^V_{{3\over 2}1}|^2+|H^A_{{3\over 2}1}|^2)\,,\nonumber\\
&&
|H^{V(A)}_{{1\over 2}0}|^2=C^\pm_{\Xi^*D^*}m_{\Omega_b}^2
\bigg[\bigg(1\mp \frac{m_{\Xi^*}}{m_{\Omega_b}}\bigg)F_1^{V(A)}
\mp\bar D^\mp_{\Xi^* D^*}\bigg(F_2^{V(A)}+\frac{m_{\Omega_b}}{m_{\Xi^*}}F_3^{V(A)}\bigg)
\mp\hat D^\pm_{\Xi^* D^*}F_4^{V(A)}\bigg]^2\,,\nonumber\\
&&
|H^{V(A)}_{{1\over 2}1}|^2=C^\pm_{\Xi^*D^*}(2m_{D^*}^2)
\bigg(F_1^{V(A)}-E^\pm_{\Xi^* D^*}F_4^{V(A)}\bigg)^2\,,\nonumber\\
&&
|H^{V(A)}_{{3\over 2}1}|^2=C^\pm_{\Xi^*D^*}(6 m_{D^*}^2)
\bigg(E^\pm_{\Xi^* D^*}F_4^{V(A)}\bigg)^2\,,
\end{eqnarray}
where
$\bar D^\pm_{{\bf B}^*M}=[(m_{\Omega_b}\pm m_{{\bf B}^*})^2-m_M^2]/(2 m_{\Omega_b}^2)$,
$\hat D^\pm_{{\bf B}^*M}=(m_{\Omega_b}^2-m_{{\bf B}^*}^2-m_M^2)/[(m_{\Omega_b}\pm m_{{\bf B}^*})^2-m_M^2]$,
and $E^\pm_{{\bf B}^* M}=(m_{\Omega_b} m_{{\bf B}^*})/[(m_{\Omega_b}\pm m_{{\bf B}^*})^2-m_M^2]$.
In Eq.~(\ref{H2}),
$(C^+_{\Xi^* D^*},C^-_{\Xi^* D^*})=(22.6,6.9)$,
$(\bar D^+_{\Xi^* D^*},\bar D^-_{\Xi^* D^*})=(0.7,0.2)$,
$(\hat D^+_{\Xi^* D^*},\hat D^-_{\Xi^* D^*})=(0.6,1.8)$, and
$(E^+_{\Xi^* D^*},E^-_{\Xi^* D^*})=(0.2,0.6)$, and $F_3^{V(A)}$ is again negligible.
We hence estimate $|H^V_{{1\over 2}0}|^2/|H^A_{{1\over 2}0}|^2
\simeq 0.04(C^+_{\Xi^*D}/C^-_{\Xi^*D})$ $\simeq 0.12$,
where 0.04 reflects that $F_1^{V}$ and $F_4^{V}$ in $|H^V_{{1\over 2}0}|^2$
substantially cancel each other, similar to the case in ${\cal B}(\Omega_b^-\to \Xi^{*-} D^0)$.
In addition to
$(|H^V_{{1\over 2}0}|^2+|H^A_{{1\over 2}0}|)^2:(|H^V_{{1\over 2}1}|^2+|H^A_{{1\over 2}1}|^2):
(|H^V_{{3\over 2}1}|^2+|H^A_{{3\over 2}1}|^2)\simeq 5:1:1$,
we obtain
\begin{eqnarray}\label{R1}
{\cal R}_2\equiv\frac{{\cal B}(\Omega_b^-\to\Xi^{*-} D^{*0})}{{\cal B}(\Omega_b^-\to\Xi^{*-}D^0)}
\simeq\frac{(m_{D^*} f_{D^*})^2 |H_{\Xi^{*} D^*}|^2}{(m_D f_{D})^2 |H_{\Xi^{*} D}|^2}\simeq 2\,,
\end{eqnarray}
where $|H_{\Xi^* D^*}|^2/|H_{\Xi^* D}|^2\simeq 1.23$.
The relations of Eqs.~(\ref{R0})~and~(\ref{R1}) are applied to
the branching fractions with $\Xi D^{(*)}(\Xi^* D^{(*)})$ as final states,
as given in Table~\ref{tab3}.
The ratio of ${\cal B}(\Omega_b^-\to\Omega^- D^{*0})$
to ${\cal B}(\Omega_b^-\to \Omega^- D^0)$ is also around 2,
where the $F_3^{V,A}$ terms are no longer negligible.
Rather, $D^-_{\Omega D}F_3^A\simeq F_4^A/2$ and
$\bar D^+_{\Omega D^*}(m_{\Omega_b}/m_{\Xi^*})F_3^A\simeq -F_4^A$
dominantly contribute to ${\cal B}(\Omega_b^-\to \Xi^{*-}D^0,\Xi^{*-}D^{*0})$,
respectively.

The ratio of ${\cal B}(\Omega_b^-\to\Omega^- J/\psi)$
to ${\cal B}(\Omega_b^-\to \Omega^- \eta_c)$,
written as
\begin{eqnarray}\label{R2}
{\cal R}_3\equiv\frac{{\cal B}(\Omega_b^-\to\Omega^- J/\psi)}{{\cal B}(\Omega_b^-\to\Omega^- \eta_c)}
\simeq\frac{(m_{J/\psi} f_{J/\psi})^2 |H_{\Omega J/\psi}|^2}
{(m_{\eta_c} f_{\eta_c})^2 |H_{\Omega \eta_c}|^2}\simeq 3.4\,,
\end{eqnarray}
presents an exceptional deviation from 2. In Eq.~(\ref{R2}),
$(m_{J/\psi} f_{J/\psi})^2/(m_{\eta_c} f_{\eta_c})^2\simeq 1.26$,
not as large as $(m_{D^*} f_{D^*})^2/(m_D f_D)^2$ in Eq.~(\ref{R1}),
is unable to account for the ratio. Nonetheless,
$(|H^{V(A)}_{{1\over 2}1}|^2,|H^{V(A)}_{{3\over 2}1}|^2)$
$\propto m_{J/\psi}^2$ play a key role in enhancing ${\cal B}(\Omega_b^-\to\Omega^- J/\psi)$.
Following Eq.~(\ref{H2}), it leads to
$(|H^V_{{1\over 2}0}|^2+|H^A_{{1\over 2}0}|)^2:(|H^V_{{1\over 2}1}|^2+|H^A_{{1\over 2}1}|^2):
(|H^V_{{3\over 2}1}|^2+|H^A_{{3\over 2}1}|^2)\simeq 6:4:4$,
which demonstrates the enhancement.
While $(|H^V_{{1\over 2}\bar 0}|^2+|H^A_{{1\over 2}\bar 0}|)^2$ is
around $1.22\times (|H^V_{{1\over 2}0}|^2+|H^A_{{1\over 2}0}|)^2$,
$|H^{V(A)}_{{1\over 2}1}|^2$ and $|H^{V(A)}_{{3\over 2}1}|^2$ raise the ratio to 3.4.

Due to the less suppressed CKM matrix elements
$V_{cb}V_{ud}^*$ in the $b\to c\bar u d$ transition, 
${\cal B}(\Omega_b^-\to\Xi^- D^{(*)0})$ and ${\cal B}(\Omega_b^-\to\Xi^{*-}D^{(*)0})$ 
can be as large as at the level of $10^{-4}$, 
together with ${\cal B}(\Omega_b^-\to\Omega^- \eta_c,\Omega^- J/\psi)$,
making them most accessible to experiments at LHCb. According to Table~\ref{tab2}, 
we present
\begin{eqnarray}\label{inW}
{\cal B}(\Omega_b^-\to\Xi^- D^0)
&=&(1.4\pm0.3, 0.012\pm 0.003, 4.1\pm 0.9)\times10^{-4}\,,\nonumber\\
{\cal B}(\Omega_b^-\to\Xi^{-}D^{*0})
&=&(3.0\pm0.9, 0.02\pm0.01, 8.4^{+2.5}_{-2.0})\times10^{-4}\,,\nonumber\\
{\cal B}(\Omega_b^-\to\Xi^{*-} D^{0})
&=&(3.1\pm 0.4, 0.03\pm 0.01, 8.7\pm 1.0)\times10^{-4}\,,\nonumber\\
{\cal B}(\Omega_b^-\to\Xi^{*-} D^{*0})
&=&(6.6\pm 1.3, 0.05\pm 0.01,18.6^{+3.7}_{-3.4})\times10^{-4}\,.
\end{eqnarray}
As the above decays proceed through the internal $W$-boson emission,
the parameter $a_2$ in Eq.~(\ref{a1a2b}) should be involved, 
resulting in three sets of the predicted values. 
Since $a_2$ is sensitive to the non-factorizable QCD corrections, 
the branching fractions can vary significantly, as demonstrated in Eq.~(\ref{inW}).

It is possible that $N_c^{eff}$ in $\Omega_b^-$ decays is close to that
in ${\bf B}_{3c}$ decays, where $N_c^{eff}=2.15\pm 0.17$
has been determined to interpret  ${\cal B}(\Lambda_b\to \Lambda J/\psi)$ and 
${\cal B}(\Xi_b^-\to \Xi^- J/\psi)$~\cite{Hsiao:2015cda,Hsiao:2015txa}. 
Nonetheless, whether $N_c^{eff}$ is close to 2 or not depends on the extractions from
more accurate experimental results, which have not been provided. In Eq.~(\ref{a1a2b}), 
since $a_1$ appears to be insensitive to the non-factorizable QCD corrections, 
corresponding decay channels can be used to avoid significant uncertainties.

With respect to $\Omega_b^-\to\Xi^- D^{(*)0},\Xi^{*-}D^{(*)0}$,
the quark-level $b\to u \bar c s$-induced decays
$\Omega_b^-\to\Xi^0 D_s^{(*)-},\Xi^{*0}D_s^{(*)-}$ 
seem to experience suppression from the factor  
$|(V_{ub}V_{cs}^*)/(V_{cb}V_{ud}^*)|^2\simeq 8\times 10^{-3}$,
but are largely compensated by $(a_1/a_2)^2$, 
with a value that can be around 20 when $N_c^{eff}=2$.
We hence obtain
\begin{eqnarray}\label{btoucs}
{\cal B}(\Omega_b^-\to \Xi^0 D_s^{-})
&=&(4.0\pm0.9, 4.5\pm1.0,5.5\pm1.2)\times10^{-5}\,,\nonumber\\
{\cal B}(\Omega_b^-\to \Xi^0 D_s^{*-})
&=&(7.8^{+2.4}_{-1.9}, 8.7^{+2.7}_{-2.1},10.9^{+3.3}_{-2.6})\times10^{-5}\,,\nonumber\\
{\cal B}(\Omega_b^-\to\Xi^{*0}D_s^{-})
&=&(7.7\pm1.0,8.7\pm1.1,10.8\pm1.3)\times10^{-5}\,,\nonumber\\
{\cal B}(\Omega_b^-\to \Xi^{*0}D_s^{*-})
&=&(15.8^{+3.3}_{-3.0},17.8^{+3.7}_{-3.4},22.1^{+4.6}_{-4.2})\times10^{-5}\,.
\end{eqnarray}
As the decays proceed through the external $W$-boson emission,
the parameter $a_1$ is involved, leading to more certain predictions.
For example, ${\cal B}(\Omega_b^-\to \Xi^0 D_s^{-})$ spans the range 
$(3.1-6.7)\times 10^{-5}$, whereas
${\cal B}(\Omega_b^-\to\Xi^- D^0)$ is predicted to be of the order $10^{-6}$ to $10^{-4}$.
Remarkably, three of the four branching fractions in Eq.~(\ref{btoucs})
are at the level of $10^{-4}$, which is beneficial for future measurements.

In summary, we have explored the singly and doubly charmful two-body $\Omega_b^-$ decays.
In the light-front quark model, we have calculated
the form factors of the $\Omega_b^-\to \Xi^{(*)}$ and $\Omega_b^-\to \Omega^-$ transitions.
We found that ${\cal B}(\Omega_b^-\to\Xi^- D^{(*)0})$ and
${\cal B}(\Omega_b^-\to\Xi^{*-}D^{(*)0})$ can be as large as at the level of $10^{-4}$,
hundred times larger than the previous studies. In our new findings,
the decays through the external $W$-boson emission lead to the branching fractions
as large as $10^{-4}$, that is,
${\cal B}(\Omega_b^-\to\Xi^0 D_s^{*-})=(5.9-14.2)\times 10^{-5}$,
${\cal B}(\Omega_b^-\to\Xi^{*0}D_s^{-})=(6.7-12.1)\times10^{-5}$, and
${\cal B}(\Omega_b^-\to\Xi^{*0}D_s^{*-})=(12.8-26.7)\times10^{-5}$,
promising to be measured by LHCb.

\section*{ACKNOWLEDGMENTS}
Y.K.H. was supported in parts by NSFC (Grants No.~12175128 and No.~11675030).
Y.L.W. was supported in part by Innovation Project of Graduate Education in Shanxi Province (2023KY430).
K.L.W. was supported in parts by NSFC (Grant No.~12205026)
and Applied Basic Research Program of Shanxi Province (Grant No.~202103021223376).

\end{document}